%
%
%
%
%
%
%
%
%
%
%
%
%
%
%
\def\standardrisposta{s }\def\reducedrisposta{r }
\def\mplarisposta{mpla }\def\zerorisposta{z }\def\bigrisposta{big }
\def\doublerisposta{d }\def\cartarisposta{e }\def\amsrisposta{y }
\newcount\ingrandimento \newcount\sinnota \newcount\dimnota
\newcount\unoduecol \newdimen\collhsize \newdimen\tothsize
\newdimen\fullhsize \newcount\controllorisposta \sinnota=1
\newskip\infralinea  \global\controllorisposta=0
\immediate\write16 { ********  Welcome to PANDA macros (Plain TeX,
AP, 1991) ******** }
\immediate\write16{ *** Edit the source file for changing the print
format ***}
%
%
%
%
%
%
%
\def\risposta{s } 
\def\srisposta{u } 
\def\arisposta{y }
%
%
%
%
%
\ifx\risposta\standardrisposta \ingrandimento=1200
\message {>> This will come out UNREDUCED << }
\dimnota=2 \unoduecol=1 \global\controllorisposta=1 \fi
\ifx\risposta\bigrisposta \ingrandimento=1440
\message {>> This will come out ENLARGED << }
\dimnota=2 \unoduecol=1 \global\controllorisposta=1 \fi
\ifx\risposta\reducedrisposta \ingrandimento=1095 \dimnota=1
\unoduecol=1  \global\controllorisposta=1
\message {>> This will come out REDUCED << } \fi
\ifx\risposta\doublerisposta \ingrandimento=1000 \dimnota=2
\unoduecol=2  \message {>> You must print this in
LANDSCAPE orientation << } \global\controllorisposta=1 \fi
\ifx\risposta\mplarisposta \ingrandimento=1000 \dimnota=1
\message {>> Mod. Phys. Lett. A format << }
\unoduecol=1 \global\controllorisposta=1 \fi
\ifx\risposta\zerorisposta \ingrandimento=1000 \dimnota=2
\message {>> Zero Magnification format << }
\unoduecol=1 \global\controllorisposta=1 \fi
\ifnum\controllorisposta=0  \ingrandimento=1200
\message {>>> ERROR IN INPUT, I ASSUME STANDARD
UNREDUCED FORMAT <<< }  \dimnota=2 \unoduecol=1 \fi
\magnification=\ingrandimento
%
%
%
%
\newdimen\eucolumnsize \newdimen\eudoublehsize \newdimen\eudoublevsize
\newdimen\uscolumnsize \newdimen\usdoublehsize \newdimen\usdoublevsize
\newdimen\eusinglehsize \newdimen\eusinglevsize \newdimen\ussinglehsize
\newskip\standardbaselineskip \newdimen\ussinglevsize
\newskip\reducedbaselineskip \newskip\doublebaselineskip
\newskip\bigbaselineskip
\eucolumnsize=12.0truecm    
\eudoublehsize=25.5truecm   
\eudoublevsize=6.5truein    
\uscolumnsize=4.4truein     
\usdoublehsize=9.4truein    
\usdoublevsize=6.8truein    
\eusinglehsize=6.3truein    
\eusinglevsize=24truecm     
\ussinglehsize=6.5truein    
\ussinglevsize=8.9truein    
\bigbaselineskip=18pt plus.2pt       
\standardbaselineskip=16pt plus.2pt  
\reducedbaselineskip=14pt plus.2pt   
\doublebaselineskip=12pt plus.2pt    
%
%
\def\Portoffset{}
\def\Landoffset{\hoffset=-.140truein}
\ifx\risposta\mplarisposta \def\Portoffset{\hoffset=1.9truecm
\voffset=1.4truecm} \fi
%
%
\def\Landspec{}
\tolerance=10000
\parskip=0pt plus2pt  \leftskip=0pt \rightskip=0pt
%
%
\ifx\risposta\bigrisposta      \infralinea=\bigbaselineskip \fi
\ifx\risposta\standardrisposta \infralinea=\standardbaselineskip \fi
\ifx\risposta\reducedrisposta  \infralinea=\reducedbaselineskip \fi
\ifx\risposta\doublerisposta   \infralinea=\doublebaselineskip \fi
\ifx\risposta\mplarisposta     \infralinea=13pt \fi
\ifx\risposta\zerorisposta     \infralinea=12pt plus.2pt\fi
\ifnum\controllorisposta=0    \infralinea=\standardbaselineskip \fi
\ifx\risposta\doublerisposta   \Landoffset \else \Portoffset \fi
\ifx\risposta\doublerisposta \ifx\srisposta\cartarisposta
\tothsize=\eudoublehsize \collhsize=\eucolumnsize
\vsize=\eudoublevsize  \else  \tothsize=\usdoublehsize
\collhsize=\uscolumnsize \vsize=\usdoublevsize \fi \else
\ifx\srisposta\cartarisposta \tothsize=\eusinglehsize
\vsize=\eusinglevsize \else  \tothsize=\ussinglehsize
\vsize=\ussinglevsize \fi \collhsize=4.4truein \fi
\ifx\risposta\mplarisposta \tothsize=5.0truein
\vsize=7.8truein \collhsize=4.4truein \fi
%
%
%
%
\newcount\contaeuler \newcount\contacyrill \newcount\contaams \newcount\contasym
\font\ninerm=cmr9  \font\eightrm=cmr8  \font\sixrm=cmr6
\font\ninei=cmmi9  \font\eighti=cmmi8  \font\sixi=cmmi6
\font\ninesy=cmsy9  \font\eightsy=cmsy8  \font\sixsy=cmsy6
\font\ninebf=cmbx9  \font\eightbf=cmbx8  \font\sixbf=cmbx6
\font\ninett=cmtt9  \font\eighttt=cmtt8  \font\nineit=cmti9
\font\eightit=cmti8 \font\ninesl=cmsl9  \font\eightsl=cmsl8
\skewchar\ninei='177 \skewchar\eighti='177 \skewchar\sixi='177
\skewchar\ninesy='60 \skewchar\eightsy='60 \skewchar\sixsy='60
\hyphenchar\ninett=-1 \hyphenchar\eighttt=-1 \hyphenchar\tentt=-1
\def\bfmath{\cmmib}                 
\font\tencmmib=cmmib10  \newfam\cmmibfam  \skewchar\tencmmib='177
\font\tencmbsy=cmbsy10  \newfam\cmbsyfam  \skewchar\tencmbsy='60
\def\scaps{\cmcsc}                 
\font\tencmcsc=cmcsc10  \newfam\cmcscfam
\ifnum\ingrandimento=1095 
 
\font\bfone=cmbx10 at 10.95pt

\font\capsone=cmcsc10 at 10.95pt 

\else  
 
\font\bfone=cmbx10 at 12pt

\font\capsone=cmcsc10 at 12pt 
\fi
\def\chapterfont#1{\xdef\ttaarr{#1}}
\def\sectionfont#1{\xdef\ppaarr{#1}}
\def\ttaarr{\bf}		
\def\ppaarr{\sl}		

%
%
%
\newfam\eufmfam \newfam\msamfam \newfam\msbmfam \newfam\eufbfam
\def\Loadeulerfonts{\global\contaeuler=1 \ifx\arisposta\amsrisposta
\font\teneufm=eufm10              
\font\eighteufm=eufm8 \font\nineeufm=eufm9 \font\sixeufm=eufm6
\font\seveneufm=eufm7  \font\fiveeufm=eufm5
\font\teneufb=eufb10              
\font\eighteufb=eufb8 \font\nineeufb=eufb9 \font\sixeufb=eufb6
\font\seveneufb=eufb7  \font\fiveeufb=eufb5
\font\teneurm=eurm10              
\font\eighteurm=eurm8 \font\nineeurm=eurm9
\font\teneurb=eurb10              
\font\eighteurb=eurb8 \font\nineeurb=eurb9
\font\teneusm=eusm10              
\font\eighteusm=eusm8 \font\nineeusm=eusm9
\font\teneusb=eusb10              
\font\eighteusb=eusb8 \font\nineeusb=eusb9
\else \def\eufm{\tt} \def\eufb{\tt} \def\eurm{\tt} \def\eurb{\tt}
\def\eusm{\tt} \def\eusb{\tt}    \fi}
\def\loadamsmath{\global\contaams=1 \ifx\arisposta\amsrisposta
\font\tenmsam=msam10 \font\ninemsam=msam9 \font\eightmsam=msam8
\font\sevenmsam=msam7 \font\sixmsam=msam6 \font\fivemsam=msam5
\font\tenmsbm=msbm10 \font\ninemsbm=msbm9 \font\eightmsbm=msbm8
\font\sevenmsbm=msbm7 \font\sixmsbm=msbm6 \font\fivemsbm=msbm5
\else \def\msbm{\bf} \fi \def\Bbb{\msbm} \def\symbl{\msam} \tenpoint}
\def\loadcyrill{\global\contacyrill=1 \ifx\arisposta\amsrisposta
\font\tenwncyr=wncyr10 \font\ninewncyr=wncyr9 \font\eightwncyr=wncyr8
\font\tenwncyb=wncyr10 \font\ninewncyb=wncyr9 \font\eightwncyb=wncyr8
\font\tenwncyi=wncyr10 \font\ninewncyi=wncyr9 \font\eightwncyi=wncyr8
\else \def\cyrill{\sl} \def\cyrilb{\sl} \def\cyrili{\sl} \fi\tenpoint}
\catcode`\@=11
\def\undefine#1{\let#1\undefined}
\def\newsymbol#1#2#3#4#5{\let\next@\relax
 \ifnum#2=\@ne\let\next@\msafam@\else
 \ifnum#2=\tw@\let\next@\msbfam@\fi\fi
 \mathchardef#1="#3\next@#4#5}
\def\mathhexbox@#1#2#3{\relax
 \ifmmode\mathpalette{}{\m@th\mathchar"#1#2#3}%
 \else\leavevmode\hbox{$\m@th\mathchar"#1#2#3$}\fi}
\def\hexnumber@#1{\ifcase#1 0\or 1\or 2\or 3\or 4\or 5\or 6\or 7\or 8\or 
9\or A\or B\or C\or D\or E\or F\fi}
\edef\msafam@{\hexnumber@\msamfam}
\edef\msbfam@{\hexnumber@\msbmfam}
\mathchardef\dabar@"0\msafam@39
\catcode`\@=12    
\def\loadamssym{\ifx\arisposta\amsrisposta  \ifnum\contaams=1 
\global\contasym=1 
\catcode`\@=11
\def\dashrightarrow{\mathrel{\dabar@\dabar@\mathchar"0\msafam@4B}}
\def\dashleftarrow{\mathrel{\mathchar"0\msafam@4C\dabar@\dabar@}}
\let\dasharrow\dashrightarrow
\def\ulcorner{\delimiter"4\msafam@70\msafam@70 }
\def\urcorner{\delimiter"5\msafam@71\msafam@71 }
\def\llcorner{\delimiter"4\msafam@78\msafam@78 }
\def\lrcorner{\delimiter"5\msafam@79\msafam@79 }
\def\yen{{\mathhexbox@\msafam@55}}
\def\checkmark{{\mathhexbox@\msafam@58 }}
\def\circledR{{\mathhexbox@\msafam@72 }}
\def\maltese{{\mathhexbox@\msafam@7A }}
\catcode`\@=12 
\input amssym.tex     \else  
\message{Panda error - First you have to use loadamsmath !!!!} \fi
\else \message{Panda error - You need the AMSFonts for these symbols 
!!!!}\fi}
\ifx\arisposta\amsrisposta
\font\sevenex=cmex7               
\font\eightex=cmex8  \font\nineex=cmex9
\font\ninecmmib=cmmib9   \font\eightcmmib=cmmib8
\font\sevencmmib=cmmib7 \font\sixcmmib=cmmib6
\font\fivecmmib=cmmib5   \skewchar\ninecmmib='177
\skewchar\eightcmmib='177  \skewchar\sevencmmib='177
\skewchar\sixcmmib='177   \skewchar\fivecmmib='177
\font\ninecmbsy=cmbsy9    \font\eightcmbsy=cmbsy8
\font\sevencmbsy=cmbsy7  \font\sixcmbsy=cmbsy6
\font\fivecmbsy=cmbsy5   \skewchar\ninecmbsy='60
\skewchar\eightcmbsy='60  \skewchar\sevencmbsy='60
\skewchar\sixcmbsy='60    \skewchar\fivecmbsy='60
\font\ninecmcsc=cmcsc9    \font\eightcmcsc=cmcsc8     \else
\def\cmmib{\fam\cmmibfam\tencmmib}\textfont\cmmibfam=\tencmmib
\scriptfont\cmmibfam=\tencmmib \scriptscriptfont\cmmibfam=\tencmmib
\def\cmbsy{\fam\cmbsyfam\tencmbsy} \textfont\cmbsyfam=\tencmbsy
\scriptfont\cmbsyfam=\tencmbsy \scriptscriptfont\cmbsyfam=\tencmbsy
\scriptfont\cmcscfam=\tencmcsc \scriptscriptfont\cmcscfam=\tencmcsc
\def\cmcsc{\fam\cmcscfam\tencmcsc} \textfont\cmcscfam=\tencmcsc \fi
\catcode`@=11
\newskip\ttglue
\gdef\tenpoint{\def\rm{\fam0\tenrm}
  \textfont0=\tenrm \scriptfont0=\sevenrm \scriptscriptfont0=\fiverm
  \textfont1=\teni \scriptfont1=\seveni \scriptscriptfont1=\fivei
  \textfont2=\tensy \scriptfont2=\sevensy \scriptscriptfont2=\fivesy
  \textfont3=\tenex \scriptfont3=\tenex \scriptscriptfont3=\tenex
  \def\mcal{\fam2 \tensy}  \def\mmit{\fam1 \teni}
  \textfont\itfam=\tenit \def\it{\fam\itfam\tenit}
  \textfont\slfam=\tensl \def\sl{\fam\slfam\tensl}
  \textfont\ttfam=\tentt \scriptfont\ttfam=\eighttt
  \scriptscriptfont\ttfam=\eighttt  \def\tt{\fam\ttfam\tentt}
  \textfont\bffam=\tenbf \scriptfont\bffam=\sevenbf
  \scriptscriptfont\bffam=\fivebf \def\bf{\fam\bffam\tenbf}
     \ifx\arisposta\amsrisposta    \ifnum\contaeuler=1
  \textfont\eufmfam=\teneufm \scriptfont\eufmfam=\seveneufm
  \scriptscriptfont\eufmfam=\fiveeufm \def\eufm{\fam\eufmfam\teneufm}
  \textfont\eufbfam=\teneufb \scriptfont\eufbfam=\seveneufb
  \scriptscriptfont\eufbfam=\fiveeufb \def\eufb{\fam\eufbfam\teneufb}
  \def\eurm{\teneurm} \def\eurb{\teneurb} \def\eusm{\teneusm}
  \def\eusb{\teneusb}    \fi    \ifnum\contaams=1
  \textfont\msamfam=\tenmsam \scriptfont\msamfam=\sevenmsam
  \scriptscriptfont\msamfam=\fivemsam \def\msam{\fam\msamfam\tenmsam}
  \textfont\msbmfam=\tenmsbm \scriptfont\msbmfam=\sevenmsbm
  \scriptscriptfont\msbmfam=\fivemsbm \def\msbm{\fam\msbmfam\tenmsbm}
     \fi      \ifnum\contacyrill=1     \def\cyrill{\tenwncyr}
  \def\cyrilb{\tenwncyb}  \def\cyrili{\tenwncyi}         \fi
  \textfont3=\tenex \scriptfont3=\sevenex \scriptscriptfont3=\sevenex
  \def\cmmib{\fam\cmmibfam\tencmmib} \scriptfont\cmmibfam=\sevencmmib
  \textfont\cmmibfam=\tencmmib  \scriptscriptfont\cmmibfam=\fivecmmib
  \def\cmbsy{\fam\cmbsyfam\tencmbsy} \scriptfont\cmbsyfam=\sevencmbsy
  \textfont\cmbsyfam=\tencmbsy  \scriptscriptfont\cmbsyfam=\fivecmbsy
  \def\cmcsc{\fam\cmcscfam\tencmcsc} \scriptfont\cmcscfam=\eightcmcsc
  \textfont\cmcscfam=\tencmcsc \scriptscriptfont\cmcscfam=\eightcmcsc
     \fi            \tt \ttglue=.5em plus.25em minus.15em
  \normalbaselineskip=12pt
  \setbox\strutbox=\hbox{\vrule height8.5pt depth3.5pt width0pt}
  \let\sc=\eightrm \let\big=\tenbig   \normalbaselines
  \baselineskip=\infralinea  \rm}
\gdef\ninepoint{\def\rm{\fam0\ninerm}
  \textfont0=\ninerm \scriptfont0=\sixrm \scriptscriptfont0=\fiverm
  \textfont1=\ninei \scriptfont1=\sixi \scriptscriptfont1=\fivei
  \textfont2=\ninesy \scriptfont2=\sixsy \scriptscriptfont2=\fivesy
  \textfont3=\tenex \scriptfont3=\tenex \scriptscriptfont3=\tenex
  \def\mcal{\fam2 \ninesy}  \def\mmit{\fam1 \ninei}
  \textfont\itfam=\nineit \def\it{\fam\itfam\nineit}
  \textfont\slfam=\ninesl \def\sl{\fam\slfam\ninesl}
  \textfont\ttfam=\ninett \scriptfont\ttfam=\eighttt
  \scriptscriptfont\ttfam=\eighttt \def\tt{\fam\ttfam\ninett}
  \textfont\bffam=\ninebf \scriptfont\bffam=\sixbf
  \scriptscriptfont\bffam=\fivebf \def\bf{\fam\bffam\ninebf}
     \ifx\arisposta\amsrisposta  \ifnum\contaeuler=1
  \textfont\eufmfam=\nineeufm \scriptfont\eufmfam=\sixeufm
  \scriptscriptfont\eufmfam=\fiveeufm \def\eufm{\fam\eufmfam\nineeufm}
  \textfont\eufbfam=\nineeufb \scriptfont\eufbfam=\sixeufb
  \scriptscriptfont\eufbfam=\fiveeufb \def\eufb{\fam\eufbfam\nineeufb}
  \def\eurm{\nineeurm} \def\eurb{\nineeurb} \def\eusm{\nineeusm}
  \def\eusb{\nineeusb}     \fi   \ifnum\contaams=1
  \textfont\msamfam=\ninemsam \scriptfont\msamfam=\sixmsam
  \scriptscriptfont\msamfam=\fivemsam \def\msam{\fam\msamfam\ninemsam}
  \textfont\msbmfam=\ninemsbm \scriptfont\msbmfam=\sixmsbm
  \scriptscriptfont\msbmfam=\fivemsbm \def\msbm{\fam\msbmfam\ninemsbm}
     \fi       \ifnum\contacyrill=1     \def\cyrill{\ninewncyr}
  \def\cyrilb{\ninewncyb}  \def\cyrili{\ninewncyi}         \fi
  \textfont3=\nineex \scriptfont3=\sevenex \scriptscriptfont3=\sevenex
  \def\cmmib{\fam\cmmibfam\ninecmmib}  \textfont\cmmibfam=\ninecmmib
  \scriptfont\cmmibfam=\sixcmmib \scriptscriptfont\cmmibfam=\fivecmmib
  \def\cmbsy{\fam\cmbsyfam\ninecmbsy}  \textfont\cmbsyfam=\ninecmbsy
  \scriptfont\cmbsyfam=\sixcmbsy \scriptscriptfont\cmbsyfam=\fivecmbsy
  \def\cmcsc{\fam\cmcscfam\ninecmcsc} \scriptfont\cmcscfam=\eightcmcsc
  \textfont\cmcscfam=\ninecmcsc \scriptscriptfont\cmcscfam=\eightcmcsc
     \fi            \tt \ttglue=.5em plus.25em minus.15em
  \normalbaselineskip=11pt
  \setbox\strutbox=\hbox{\vrule height8pt depth3pt width0pt}
  \let\sc=\sevenrm \let\big=\ninebig \normalbaselines\rm}
\gdef\eightpoint{\def\rm{\fam0\eightrm}
  \textfont0=\eightrm \scriptfont0=\sixrm \scriptscriptfont0=\fiverm
  \textfont1=\eighti \scriptfont1=\sixi \scriptscriptfont1=\fivei
  \textfont2=\eightsy \scriptfont2=\sixsy \scriptscriptfont2=\fivesy
  \textfont3=\tenex \scriptfont3=\tenex \scriptscriptfont3=\tenex
  \def\mcal{\fam2 \eightsy}  \def\mmit{\fam1 \eighti}
  \textfont\itfam=\eightit \def\it{\fam\itfam\eightit}
  \textfont\slfam=\eightsl \def\sl{\fam\slfam\eightsl}
  \textfont\ttfam=\eighttt \scriptfont\ttfam=\eighttt
  \scriptscriptfont\ttfam=\eighttt \def\tt{\fam\ttfam\eighttt}
  \textfont\bffam=\eightbf \scriptfont\bffam=\sixbf
  \scriptscriptfont\bffam=\fivebf \def\bf{\fam\bffam\eightbf}
     \ifx\arisposta\amsrisposta   \ifnum\contaeuler=1
  \textfont\eufmfam=\eighteufm \scriptfont\eufmfam=\sixeufm
  \scriptscriptfont\eufmfam=\fiveeufm \def\eufm{\fam\eufmfam\eighteufm}
  \textfont\eufbfam=\eighteufb \scriptfont\eufbfam=\sixeufb
  \scriptscriptfont\eufbfam=\fiveeufb \def\eufb{\fam\eufbfam\eighteufb}
  \def\eurm{\eighteurm} \def\eurb{\eighteurb} \def\eusm{\eighteusm}
  \def\eusb{\eighteusb}       \fi    \ifnum\contaams=1
  \textfont\msamfam=\eightmsam \scriptfont\msamfam=\sixmsam
  \scriptscriptfont\msamfam=\fivemsam \def\msam{\fam\msamfam\eightmsam}
  \textfont\msbmfam=\eightmsbm \scriptfont\msbmfam=\sixmsbm
  \scriptscriptfont\msbmfam=\fivemsbm \def\msbm{\fam\msbmfam\eightmsbm}
     \fi       \ifnum\contacyrill=1     \def\cyrill{\eightwncyr}
  \def\cyrilb{\eightwncyb}  \def\cyrili{\eightwncyi}         \fi
  \textfont3=\eightex \scriptfont3=\sevenex \scriptscriptfont3=\sevenex
  \def\cmmib{\fam\cmmibfam\eightcmmib}  \textfont\cmmibfam=\eightcmmib
  \scriptfont\cmmibfam=\sixcmmib \scriptscriptfont\cmmibfam=\fivecmmib
  \def\cmbsy{\fam\cmbsyfam\eightcmbsy}  \textfont\cmbsyfam=\eightcmbsy
  \scriptfont\cmbsyfam=\sixcmbsy \scriptscriptfont\cmbsyfam=\fivecmbsy
  \def\cmcsc{\fam\cmcscfam\eightcmcsc} \scriptfont\cmcscfam=\eightcmcsc
  \textfont\cmcscfam=\eightcmcsc \scriptscriptfont\cmcscfam=\eightcmcsc
     \fi             \tt \ttglue=.5em plus.25em minus.15em
  \normalbaselineskip=9pt
  \setbox\strutbox=\hbox{\vrule height7pt depth2pt width0pt}
  \let\sc=\sixrm \let\big=\eightbig \normalbaselines\rm }
\gdef\tenbig#1{{\hbox{$\left#1\vbox to8.5pt{}\right.\n@space$}}}
\gdef\ninebig#1{{\hbox{$\textfont0=\tenrm\textfont2=\tensy
   \left#1\vbox to7.25pt{}\right.\n@space$}}}
\gdef\eightbig#1{{\hbox{$\textfont0=\ninerm\textfont2=\ninesy
   \left#1\vbox to6.5pt{}\right.\n@space$}}}
\def\alternativefont#1#2{\ifx\arisposta\amsrisposta \relax \else
\xdef#1{#2} \fi}
\global\contaeuler=0 \global\contacyrill=0 \global\contaams=0
%
%
%
%
\newbox\fotlinebb \newbox\hedlinebb \newbox\leftcolumn
\gdef\makeheadline{\vbox to 0pt{\vskip-22.5pt
     \fullline{\vbox to8.5pt{}\the\headline}\vss}\nointerlineskip}
\gdef\makehedlinebb{\vbox to 0pt{\vskip-22.5pt
     \fullline{\vbox to8.5pt{}\copy\hedlinebb\hfil
     \line{\hfill\the\headline\hfill}}\vss} \nointerlineskip}
\gdef\makefootline{\baselineskip=24pt \fullline{\the\footline}}
\gdef\makefotlinebb{\baselineskip=24pt
    \fullline{\copy\fotlinebb\hfil\line{\hfill\the\footline\hfill}}}
\gdef\doubleformat{\shipout\vbox{\Landspec\makehedlinebb
     \fullline{\box\leftcolumn\hfil\columnbox}\makefotlinebb}
     \advancepageno}
\gdef\columnbox{\leftline{\pagebody}}
\gdef\line#1{\hbox to\hsize{\hskip\leftskip#1\hskip\rightskip}}
\gdef\fullline#1{\hbox to\fullhsize{\hskip\leftskip{#1}%
\hskip\rightskip}}
\gdef\footnote#1{\let\@sf=\empty
         \ifhmode\edef\#sf{\spacefactor=\the\spacefactor}\/\fi
         #1\@sf\vfootnote{#1}}
\gdef\vfootnote#1{\insert\footins\bgroup
         \ifnum\dimnota=1  \eightpoint\fi
         \ifnum\dimnota=2  \ninepoint\fi
         \ifnum\dimnota=0  \tenpoint\fi
         \interlinepenalty=\interfootnotelinepenalty
         \splittopskip=\ht\strutbox
         \splitmaxdepth=\dp\strutbox \floatingpenalty=20000
         \leftskip=\oldssposta \rightskip=\olddsposta
         \spaceskip=0pt \xspaceskip=0pt
         \ifnum\sinnota=0   \textindent{#1}\fi
         \ifnum\sinnota=1   \item{#1}\fi
         \footstrut\futurelet\next\fo@t}
\gdef\fo@t{\ifcat\bgroup\noexpand\next \let\next\f@@t
             \else\let\next\f@t\fi \next}
\gdef\f@@t{\bgroup\aftergroup\@foot\let\next}
\gdef\f@t#1{#1\@foot} \gdef\@foot{\strut\egroup}
\gdef\footstrut{\vbox to\splittopskip{}}
\skip\footins=\bigskipamount
\count\footins=1000  \dimen\footins=8in
\catcode`@=12
\tenpoint
\ifnum\unoduecol=1 \hsize=\tothsize   \fullhsize=\tothsize \fi
\ifnum\unoduecol=2 \hsize=\collhsize  \fullhsize=\tothsize \fi
\global\let\lrcol=L      \ifnum\unoduecol=1
\output{\plainoutput{\ifnum\tipbnota=2 \clearnmbnota\fi}} \fi
\ifnum\unoduecol=2 \output{\if L\lrcol
     \global\setbox\leftcolumn=\columnbox
     \global\setbox\fotlinebb=\line{\hfill\the\footline\hfill}
     \global\setbox\hedlinebb=\line{\hfill\the\headline\hfill}
     \advancepageno  \global\let\lrcol=R
     \else  \doubleformat \global\let\lrcol=L \fi
     \ifnum\outputpenalty>-20000 \else\dosupereject\fi
     \ifnum\tipbnota=2\clearnmbnota\fi }\fi
\def\ifdoublepage{\ifnum\unoduecol=2 }
\gdef\yespagenumbers{\footline={\hss\tenrm\folio\hss}}
\gdef\ciao{ \ifnum\fdefcontre=1 \endfdef\fi
     \par\vfill\supereject \ifnum\unoduecol=2
     \if R\lrcol  \headline={}\nopagenumbers\null\vfill\eject
     \fi\fi \end}

\newskip\olddsposta \newskip\oldssposta
\global\oldssposta=\leftskip \global\olddsposta=\rightskip

\def\filldots{\leaders\hbox to 1em{\hss.\hss}\hfill}
\def\inquadrb#1 {\vbox {\hrule  \hbox{\vrule \vbox {\vskip .2cm
    \hbox {\ #1\ } \vskip .2cm } \vrule  }  \hrule} }
 \def\newline{\hfil\break}
\def\jump{\vskip\baselineskip} \newskip\iinnffrr
\def\sjump{\iinnffrr=\baselineskip
          \divide\iinnffrr by 2 \vskip\iinnffrr}
\def\bjump{\vskip\baselineskip \vskip\baselineskip}
\newcount\nmbnota  \def\clearnmbnota{\global\nmbnota=0}
\newcount\tipbnota \def\letterfootnote{\global\tipbnota=1}

\def\note#1{\global\advance\nmbnota by 1 \ifnum\tipbnota=1
    \footnote{$^{\rm\nttlett}$}{#1} \else {\ifnum\tipbnota=2
    \footnote{$^{\nttsymb}$}{#1}
    \else\footnote{$^{\the\nmbnota}$}{#1}\fi}\fi}
\def\nttlett{\ifcase\nmbnota \or a\or b\or c\or d\or e\or f\or
g\or h\or i\or j\or k\or l\or m\or n\or o\or p\or q\or r\or
s\or t\or u\or v\or w\or y\or x\or z\fi}
\def\nttsymb{\ifcase\nmbnota \or\dag\or\sharp\or\ddag\or\star\or
\natural\or\flat\or\clubsuit\or\diamondsuit\or\heartsuit
\or\spadesuit\fi}   \clearnmbnota
\def\numberfootnote{\global\tipbnota=0} \numberfootnote
\def\setnote#1{\expandafter\xdef\csname#1\endcsname{
\ifnum\tipbnota=1 {\rm\nttlett} \else {\ifnum\tipbnota=2
{\nttsymb} \else \the\nmbnota\fi}\fi} }
\newcount\nbmfig  \def\clearnbmfig{\global\nbmfig=0}
\gdef\figure{\global\advance\nbmfig by 1
      {\rm fig. \the\nbmfig}}   \clearnbmfig
\def\setfig#1{\expandafter\xdef\csname#1\endcsname{fig. \the\nbmfig}}
 \def\endformula{\eqno\numero $$}
 \def\efr{\endformula}
\newcount\frmcount \def\clearfrmcount{\global\frmcount=0}
\def\numero{\global\advance\frmcount by 1   \ifnum\indappcount=0
  {\ifnum\cpcount <1 {\hbox{\rm (\the\frmcount )}}  \else
  {\hbox{\rm (\the\cpcount .\the\frmcount )}} \fi}  \else
  {\hbox{\rm (\applett .\the\frmcount )}} \fi}
\def\nfr{\nameformula}    \def\numali{\numero}
\def\nameformula#1{\global\advance\frmcount by 1%
{\ifnum\indappcount=0%
{\ifnum\cpcount<1\xdef\spzzttrra{(\the\frmcount )}%
\else\xdef\spzzttrra{(\the\cpcount .\the\frmcount )}\fi}%
\else\xdef\spzzttrra{(\applett .\the\frmcount )}\fi}%
\expandafter\xdef\csname#1\endcsname{\spzzttrra}%
\eqno{\ifnum\draftnum=0\hbox{\rm\spzzttrra}\else%
\hbox{$\buildchar{\rm\spzzttrra}{\tt\scriptscriptstyle#1}{}$}\fi}$$}
\def\nameali#1{\global\advance\frmcount by 1%
{\ifnum\indappcount=0%
{\ifnum\cpcount<1\xdef\spzzttrra{(\the\frmcount )}%
\else\xdef\spzzttrra{(\the\cpcount .\the\frmcount )}\fi}%
\else\xdef\spzzttrra{(\applett .\the\frmcount )}\fi}%
\expandafter\xdef\csname#1\endcsname{\spzzttrra}%
\ifnum\draftnum=0\hbox{\rm\spzzttrra}\else%
\hbox{$\buildchar{\rm\spzzttrra}{\tt\scriptscriptstyle#1}{}$}\fi}
\clearfrmcount
\newcount\cpcount \def\clearcpcount{\global\cpcount=0}
\newcount\subcpcount \def\clearsubcpcount{\global\subcpcount=0}
\newcount\appcount \def\clearappcount{\global\appcount=0}
\newcount\indappcount \def\clearindappcount{\indappcount=0}
\newcount\sottoparcount 

\def\applett{\ifcase\appcount  \or {A}\or {B}\or {C}\or
{D}\or {E}\or {F}\or {G}\or {H}\or {I}\or {J}\or {K}\or {L}\or
{M}\or {N}\or {O}\or {P}\or {Q}\or {R}\or {S}\or {T}\or {U}\or
{V}\or {W}\or {X}\or {Y}\or {Z}\fi    \ifnum\appcount<0
\immediate\write16 {Panda ERROR - Appendix: counter "appcount"
out of range}\fi  \ifnum\appcount>26  \immediate\write16 {Panda
ERROR - Appendix: counter "appcount" out of range}\fi}
\clearappcount  \clearindappcount \newcount\connttrre
\def\clearconnttrre{\global\connttrre=0} \newcount\countref
\def\clearcountref{\global\countref=0} \clearcountref
\def\chapter#1{\global\advance\cpcount by 1 \clearfrmcount
                 \goodbreak\null\vbox{\jump\nobreak
                 \clearsubcpcount\clearindappcount
                 \itemitem{\ttaarr\the\cpcount .\qquad}{\ttaarr #1}
                 \par\nobreak\jump\sjump}\nobreak}
\def\section#1{\global\advance\subcpcount by 1 \goodbreak\null
               \vbox{\sjump\nobreak\ifnum\indappcount=0
                 {\ifnum\cpcount=0 {\itemitem{\ppaarr
               .\the\subcpcount\quad\enskip\ }{\ppaarr #1}\par} \else
                 {\itemitem{\ppaarr\the\cpcount .\the\subcpcount\quad
                  \enskip\ }{\ppaarr #1} \par}  \fi}
                \else{\itemitem{\ppaarr\applett .\the\subcpcount\quad
                 \enskip\ }{\ppaarr #1}\par}\fi\nobreak\jump}\nobreak}
\clearsubcpcount
\def\appendix#1{\global\advance\appcount by 1 \clearfrmcount
                  \goodbreak\null\vbox{\jump\nobreak
                  \global\advance\indappcount by 1 \clearsubcpcount
          \itemitem{ }{\hskip-40pt\ttaarr Appendix\ \applett :\ #1}
             \nobreak\jump\sjump}\nobreak}
\clearappcount \clearindappcount
\def\references{\goodbreak\null\vbox{\jump\nobreak
   \itemitem{}{\ttaarr References} \nobreak\jump\sjump}\nobreak}

\def\introsumm{\clearindappcount\clearappcount\clearcpcount
                  \clearsubcpcount\goodbreak\null\vbox{\jump\nobreak
  \itemitem{}{\ttaarr Introduction and Summary} \nobreak\jump\sjump}\nobreak}
\clearcpcount\clearcountref

\def\setchap#1{\ifnum\indappcount=0{\ifnum\subcpcount=0%
\xdef\spzzttrra{\the\cpcount}%
\else\xdef\spzzttrra{\the\cpcount .\the\subcpcount}\fi}
\else{\ifnum\subcpcount=0 \xdef\spzzttrra{\applett}%
\else\xdef\spzzttrra{\applett .\the\subcpcount}\fi}\fi
\expandafter\xdef\csname#1\endcsname{\spzzttrra}}
\newcount\draftnum \newcount\ppora   \newcount\ppminuti
\global\ppora=\time   \global\ppminuti=\time
\global\divide\ppora by 60  \draftnum=\ppora
\multiply\draftnum by 60    \global\advance\ppminuti by -\draftnum
\def\droggi{\number\day /\number\month /\number\year\ \the\ppora
:\the\ppminuti}     \global\draftnum=0
\def\draftcomment#1{\ifnum\draftnum=0 \relax \else
{\ {\bf ***}\ #1\ {\bf ***}\ }\fi} 
%
%
\catcode`@=11
\gdef\Ref#1{\expandafter\ifx\csname @rrxx@#1\endcsname\relax%
{\global\advance\countref by 1    \ifnum\countref>200
\immediate\write16 {Panda ERROR - Ref: maximum number of references
exceeded}  \expandafter\xdef\csname @rrxx@#1\endcsname{0}\else
\expandafter\xdef\csname @rrxx@#1\endcsname{\the\countref}\fi}\fi
\ifnum\draftnum=0 \csname @rrxx@#1\endcsname \else#1\fi}
\gdef\beginref{\ifnum\draftnum=0  \gdef\Rref{\fairef}
\gdef\endref{\scriviref} \else\relax\fi
\ifx\risposta\mplarisposta \ninepoint \fi
\baselineskip=12pt \parskip 2pt plus.2pt }
\def\Reflab#1{[#1]} \gdef\Rref#1#2{\item{\Reflab{#1}}{#2}}
\gdef\endref{\relax}  \newcount\conttemp
\gdef\fairef#1#2{\expandafter\ifx\csname @rrxx@#1\endcsname\relax
{\global\conttemp=0 \immediate\write16 {Panda ERROR - Ref: reference
[#1] undefined}} \else
{\global\conttemp=\csname @rrxx@#1\endcsname } \fi
\global\advance\conttemp by 50  \global\setbox\conttemp=\hbox{#2} }
\gdef\scriviref{\clearconnttrre\conttemp=50
\loop\ifnum\connttrre<\countref \advance\conttemp by 1
\advance\connttrre by 1
\item{\Reflab{\the\connttrre}}{\unhcopy\conttemp} \repeat}
\clearcountref \clearconnttrre
\catcode`@=12
\ifx\risposta\mplarisposta \def\Reflab#1{#1.} \letterfootnote \fi
%
%

\def\slashchar#1{\setbox0=\hbox{$#1$} \dimen0=\wd0
     \setbox1=\hbox{/} \dimen1=\wd1 \ifdim\dimen0>\dimen1
      \rlap{\hbox to \dimen0{\hfil/\hfil}} #1 \else
      \rlap{\hbox to \dimen1{\hfil$#1$\hfil}} / \fi}
\ifx\oldchi\undefined \let\oldchi=\chi
  \def\cchi{{\raise 1pt\hbox{$\oldchi$}}} \let\chi=\cchi \fi
\ifnum\contasym=1 \else \fi

\def\frac#1#2{{\textstyle{#1 \over #2}}}

\def\half{\ifinner {\scriptstyle {1 \over 2}}\else {1 \over 2} \fi}

\def\simge{\rlap{\raise 2pt \hbox{$>$}}{\lower 2pt \hbox{$\sim$}}}
\def\simle{\rlap{\raise 2pt \hbox{$<$}}{\lower 2pt \hbox{$\sim$}}}

\def\buildchar#1#2#3{{\null\!\mathop{#1}\limits^{#2}_{#3}\!\null}}

\def\vbig#1#2{{\vbigd@men=#2\divide\vbigd@men by 2%
\hbox{$\left#1\vbox to \vbigd@men{}\right.\n@space$}}}

\def\noblackbox{\overfullrule=0pt} 
%
%
\newcount\fdefcontre \newcount\fdefcount \newcount\indcount
\newread\filefdef  \newread\fileftmp  \newwrite\filefdef
\newwrite\fileftmp     \def\strip #1*.A {#1}%
\def\futuredef#1{\beginfdef
\expandafter\ifx\csname#1\endcsname\relax%
{\immediate\write\fileftmp{#1*.A}%
\immediate\write16 {Panda Warning - fdef: macro "#1" on page
\the\pageno \space undefined}
\ifnum\draftnum=0 \expandafter\xdef\csname#1\endcsname{(?)}
\else \expandafter\xdef\csname#1\endcsname{(#1)}\fi
\global\advance\fdefcount by 1}\fi\csname#1\endcsname}

\def\beginfdef{\ifnum\fdefcontre=0
\immediate\openin\filefdef\jobname.fdef
\immediate\openout\fileftmp\jobname.ftmp
\global\fdefcontre=1  \ifeof\filefdef \immediate\write16 {Panda
WARNING - fdef: file \jobname.fdef not found, run TeX again}
\else \immediate\read\filefdef to\spzzttrra
\global\advance\fdefcount by \spzzttrra
\indcount=0 \loop\ifnum\indcount<\fdefcount
\advance\indcount by 1%
\immediate\read\filefdef to\spezttrra%
\immediate\read\filefdef to\sppzttrra%
\edef\spzzttrra{\expandafter\strip\spezttrra}%
\immediate\write\fileftmp {\spzzttrra *.A}
\expandafter\xdef\csname\spzzttrra\endcsname{\sppzttrra}%
\repeat \fi \immediate\closein\filefdef \fi}
\def\endfdef{\immediate\closeout\fileftmp   \ifnum\fdefcount>0
\immediate\openin\fileftmp \jobname.ftmp
\immediate\openout\filefdef \jobname.fdef
\immediate\write\filefdef {\the\fdefcount}   \indcount=0
\loop\ifnum\indcount<\fdefcount    \advance\indcount by 1
\immediate\read\fileftmp to\spezttrra
\edef\spzzttrra{\expandafter\strip\spezttrra}
\immediate\write\filefdef{\spzzttrra *.A}
\edef\spezttrra{\string{\csname\spzzttrra\endcsname\string}}
\iwritel\filefdef{\spezttrra}
\repeat  \immediate\closein\fileftmp \immediate\closeout\filefdef
\immediate\write16 {Panda Warning - fdef: Label(s) may have changed,
re-run TeX to get them right}\fi}
\def\iwritel#1#2{\newlinechar=-1
{\newlinechar=`\ \immediate\write#1{#2}}\newlinechar=-1}
\global\fdefcontre=0 \global\fdefcount=0 \global\indcount=0
%
%
%
\mathchardef\alpha="710B   \mathchardef\beta="710C
\mathchardef\gamma="710D   \mathchardef\delta="710E
\mathchardef\epsilon="710F   \mathchardef\zeta="7110
\mathchardef\eta="7111   \mathchardef\theta="7112
\mathchardef\iota="7113   \mathchardef\kappa="7114
\mathchardef\lambda="7115   \mathchardef\mu="7116
\mathchardef\nu="7117   \mathchardef\xi="7118
\mathchardef\pi="7119   \mathchardef\rho="711A
\mathchardef\sigma="711B   \mathchardef\tau="711C
\mathchardef\upsilon="711D   \mathchardef\phi="711E
\mathchardef\chi="711F   \mathchardef\psi="7120
\mathchardef\omega="7121   \mathchardef\varepsilon="7122
\mathchardef\vartheta="7123   \mathchardef\varpi="7124
\mathchardef\varrho="7125   \mathchardef\varsigma="7126
\mathchardef\varphi="7127
%
%
\null
%
%
%
%
%
\loadamsmath 
\chapterfont{\bfone} \sectionfont{\scaps}
\noblackbox

\def\cptarrow{\ \buildchar{\longrightarrow}{{\rm{\scriptscriptstyle 
              CPT}}}{ }\ }
\def\wcptarrow{\ \buildchar{\longrightarrow}{{\rm{\scriptscriptstyle 
              WS-CPT}}}{ }\ }
\def\srarrow{\ \buildchar{\longrightarrow}{{\rm{\scriptscriptstyle 
              SR}}}{ }\ }
\def\bpzarrow{\ \buildchar{\longrightarrow}{{\rm{\scriptscriptstyle 
              BPZ}}}{ }\ }
\def\eqmodone{\ \buildchar{=}{{\rm{\scriptscriptstyle MOD\ 1}}}{ }\ }
\def\eqmodtwo{\ \buildchar{=}{{\rm{\scriptscriptstyle MOD\ 2}}}{ }\ }
\def\eqope{\ \buildchar{=}{\rm\scriptscriptstyle OPE}{ }\ }
\def\eqtope{\ \buildchar{\sim}{\rm\scriptscriptstyle OPE}{ }\ }
\def\eqcpt{\ \buildchar{=}{\rm\scriptscriptstyle CPT}{ }\ }

\def\wew#1{\langle\langle\, #1\, \rangle\rangle}
\def\di{{\rm d}}
\nopagenumbers
{\baselineskip=12pt
\line{\hfill IFUM-523/FT}
\line{\hfill NBI-HE-96-03}
\line{\hfill hep-th/9602026}
\line{\hfill January, 1996}}
{\baselineskip=14pt
\vfill
\centerline{\capsone CPT Invariance of String Models}
\sjump
\centerline{\capsone in a Minkowski Background}
\bjump
\centerline{\scaps Andrea Pasquinucci~\footnote{$^\dagger$}{Supported by 
a MURST Fellowship.}}
\sjump
\centerline{\sl Dipartimento di Fisica, Universit\`a di Milano,}
\centerline{\sl via Celoria 16, I-20133 Milano, Italy}
\jump
\centerline{\scaps Kaj Roland~\footnote{$^\ddagger$}{Supported by the 
Carlsberg Foundation.}}
\sjump
\centerline{\sl The Niels Bohr Institute, University of Copenhagen,}
\centerline{\sl Blegdamsvej 17, DK-2100, Copenhagen, Denmark}
\bjump \vfill
\centerline{\capsone ABSTRACT}
\sjump
\noindent
We study the space-time CPT properties of string theories formulated in 
a flat Minkowski background of even dimension.  We define CPT as a 
world-sheet transformation acting on the vertex operators and we prove 
the CPT invariance of the string $S$-matrix elements. 
Some related issues, including the connection between spin and 
statistics of physical string states, are also considered.

\sjump \vfill 
%
\pageno=0 \eject }
\yespagenumbers\pageno=1
%
\null\bjump

\introsumm
Recently there has been some interest in the question of possible
CPT non-conservation 
in string theory. Indeed, some mechanisms have been proposed that would
lead to CPT-breaking effects that might be detected in the next generation of 
experiments~[\Ref{Pott},\Ref{Ellis}].  If observed, these effects
could give the first experimental evidence of the existence or non-existence
of strings.

Accordingly the issue of CPT invariance in string theory
is quite interesting both from a theoretical and an
experimental point of view. From a theoretical point of view, not too much
is known and published on the space-time CPT properties of string theory. 
Sonoda [\Ref{Sonoda}] discussed and proved the space-time CPT 
theorem at the level of string perturbation theory for ten-dimensional 
heterotic strings in a Minkowski background. 
Kostelecky and Potting [\Ref{Pott}] proved the {\it dynamical}
CPT invariance of the open bosonic and super string field theories, 
formulated in flat backgrounds --but they also suggested a method 
whereby CPT might be broken {\it spontaneously}, based on the possibility 
of a CPT non-invariant ground state.

For more complicated string models no general analysis has appeared.
With this paper, we would therefore like to start a detailed 
investigation of the space-time CPT properties of general
string theories. Should we always
expect some kind of space-time CPT theorem to hold?

Let us start by reviewing the situation in field theory.
The CPT theorem 
[\Ref{Pauli},\Ref{Luders},\Ref{SW},\Ref{others}]~\note{For a general 
discussion of the CPT theorem in field theory see for example
ref. [\Ref{Weinbook}].} asserts that any 
quantum field theory is invariant under CPT transformations assuming 
that it satisfies the following very mild assumptions~[\Ref{SW}]:

\itemitem{a)}{Lorentz invariance,}

\itemitem{b)}{The energy is positive definite and there exists a 
Poincar\'{e}-invariant vacuum, unique up to a phase factor,}

\itemitem{c)}{Local commutativity, i.e. field operators at space-like 
separations either commute or anti-commute.}

\noindent
These assumptions also imply the spin-statistics theorem, i.e. fields of 
integer (half odd integer) spin~\note{In $D$ dimensions the concept of 
integer (half odd integer) spin is replaced by the concept of 
representations of the little group 
with integer (half odd integer) weights. We 
always use the word ``spin'' for convenience.} are 
quantized with respect to Bose (Fermi) statistics.

The proof of the CPT theorem in field theory (see ref. [\Ref{SW}]) is 
actually based only on the following two assumptions:

\itemitem{a)}{Lorentz invariance,}

\itemitem{d)}{spin-statistics theorem.}

\sjump

Turning our attention to string theory~[\Ref{GSW}], 
we immediately see that for
a generic string theory, formulated in a curved space-time background,  
none of the assumptions a), b) or c) is actually satisfied:
There is no global concept of Lorentz invariance, potential energy in a 
gravitational field can be negative and, since strings are extended objects, 
local commutativity is not satisfied 
either~[\Ref{elienonloc},\Ref{martcaus},\Ref{lowenonloc}]. The latter 
problem persists even if we restrict ourselves to a flat background.
Even so, CPT invariance may still have a role to play in quantum 
gravity, as pointed out by Hawking~[\Ref{Hawking}], and also in string 
theory.

Clearly the first step consists in trying to define what the CPT 
transformation should be in a generic string theory. 
For non-trivial backgrounds this is likely to be a difficult problem. 
After all, even in a quantum field theory the CPT invariance
$\phi(x) \rightarrow \phi(-x)$ can be quite obscure if we formulate the 
theory in a non-trivial background. Then the CPT symmetry 
involves not just a transformation of the quantum field but also 
a change of the background. 

Since two-dimensional conformal field theory (CFT) is the unifying 
framework common to string theories in all backgrounds, the most 
promising approach would seem to be a world-sheet formulation of the
space-time CPT transformation, where in a given string model
the CPT transformation acts on the fields and states of the underlying
CFT.

After having made an adequate definition of the CPT transformation
as a world-sheet map, 
one should then make a precise formulation of the requirement of CPT 
invariance and proceed to look either for a proof of CPT invariance, 
or find a counter-example.

For a generic string theory one can then think of at least three possible 
situations: the CPT theorem holds true, it holds true only perturbatively, 
it doesn't hold at all. More complicated cases can be imagined, but as a 
starting point we can limit ourselves to these three simple cases.

In this paper we will analyze the simplest class of models, that of 
first-quantized string models in an even-dimensional Minkowski 
background. 
Since for first-quantized string theory we do not have an off-shell 
(lagrangian) formulation we need to formulate and 
prove the CPT theorem at the level of the $S$-matrix elements. 
Since the whole scattering theory does not make much physical sense 
unless the $S$-matrix is unitary, we will always assume this to be the 
case for the string theories we consider.

Our strategy will be the following. 
We first define the space-time CPT transformation. It is 
a {\sl world-sheet\/} transformation acting on the world-sheet operators 
which are the building blocks of the vertex operators. 

\noindent
Then, using the hypothesis of:
\item{$\bullet$}{explicit Lorentz invariance of the scattering 
amplitudes,}
\item{$\bullet$}{validity of the spin-statistics theorem for the 
physical space-time spectrum,}

\noindent
we will formally prove that every $S$-matrix element in these models is 
invariant under the space-time CPT transformation at any loop order. 

In the proof we will not need to evaluate explicitly any $S$-matrix 
element. It will be enough to know that 
the $S$-matrix elements are given by the well-known 
operator formulation of the Polyakov path integral; 
and that the world-sheet correlation functions 
of the physical state vertex operators satisfy the twin requirements of 
Lorentz invariance and spin-statistics. For this reason our results are 
applicable to both bosonic and type II superstring theories, as well as 
heterotic ones.

{}From this point of view, our result can even be considered to be 
{\sl non-perturbative\/} as far as one is able to give a non perturbative 
definition of the Polyakov path integral, including the summation over 
topologies, which preserves the three assumptions listed above.

Obviously, since we have defined CPT as a transformation on the 
world-sheet, we need to verify that, apart from being an invariance of 
the $S$-matrix, it also has the correct space-time behaviour. 
We show that the CPT transformation is indeed anti-linear and transforms 
the vertex operator of any physical 
string state into a vertex operator also describing a physical string state, 
but with opposite spin
and having the opposite sign of all gauge charges. 

In this way, the CPT transformation introduces
in a natural way the concept of space-time ``anti-particle'' in a string
theory, by giving a precise world-sheet map which relates the 
``particle'' 
string-state of given helicity to the ``anti-particle'' string state of 
opposite helicity.

We give some explicit examples of this in the framework of 
four-dimensional string models
built with free world-sheet fermions [\Ref{KLT},\Ref{Anto},\Ref{Bluhm}], 
henceforth referred to as KLT models, since the formulation we use is 
that of Kawai, Lewellen and Tye [\Ref{KLT}].

Since our proof of CPT invariance is based on 
Lorentz invariance and spin-statistics, it becomes of interest to 
consider more carefully the role of the spin-statistics theorem 
in string theory. In field theory it is exactly in the proof 
of the spin-statistics theorem [\Ref{otherSS},\Ref{SW},\Ref{DHR}] 
that the assumption c) of local 
commutativity enters. In string theory, local commutativity does not really 
make sense at the first quantized level, but it does make sense at the 
level of string field theory, where it was found [\Ref{lowenonloc}] that 
the commutator of string fields is in general non-vanishing outside the 
string light cone. With local commutativity violated in string theory,
should we expect the spin-statistics relation to hold? 

In the Neveu-Schwarz Ramond formulation of superstring theory, 
the spin-statistics theorem is closely related to 
the GSO projection [\Ref{GSO},\Ref{seiberg-witten},\Ref{Anto}]. 
It is due to the GSO projection that
string states of integer space-time spin contribute to the 
partition function with an overall plus sign, whereas string states of 
half odd integer space-time spin contribute to the partition function
with an overall minus sign. Actually, what we need for the proof of the CPT 
theorem is a seemingly stronger statement of spin-statistics, 
namely that any two vertex operators describing 
physical states of half odd integer space-time spin anti-commute whereas 
any other pair of physical vertex operators commute. Even if
GSO by itself is not enough to guarantee this ``canonical'' 
spin-statistics relation, our expectation is 
that the GSO conditions are sufficient to guarantee that, by the 
introduction of appropriate cocycle operators, 
one can build vertex operators which do satisfy the ``canonical''
spin-statistics relation.
However, we do not have any general proof of this conjecture.

As an example, we discuss the situation for a very simple (bosonized) 
KLT model where we find that it is indeed possible to make a cocycle 
choice so that the vertex operators satisfy the ``canonical'' spin-statistics 
relations. We also notice that there do exist other choices of cocycles, 
which lead to vertex operators not satisfying the ``canonical'' 
spin-statistics relations. Accordingly, these choices lead to
theories with para-statistics, to use the terminolgy of ref.~[\Ref{SW}],
which at the end should be physically equivalent to the theories with 
``canonical'' statistics.

The paper is organized as follows.

In section 1 we briefly review the CPT theorem in quantum field
theory in various dimensions, including the special case of 
two-dimensional conformal field theories.

In section 2 we consider string theories in an even-dimensional 
Minkowski background and define the space-time CPT transformation
as a world-sheet transformation acting on the vertex operators of 
physical string states. We then prove that our definition leads to 
CPT invariance of the $S$-matrix elements. 

In section 3 we verify that the CPT transformation we have defined  
does indeed have the correct space-time properties when acting on 
physical string states, i.e. that it is an anti-linear map which
changes sign on the spin and on all charges,
something we further illustrate by means of two examples. 
We also comment on the relation between the {\sl space-time\/} and 
{\sl world-sheet\/} CPT transformations in string theory. 

In section 4 we discuss various aspects of the spin-statistics theorem 
in string theories.

Finally, section 5 contains our concluding remarks and some open problems. 

In the Appendix we give the definition of the helicity operator in 
four-dimensional string theory, illustrated by means of an example.

\chapter{The CPT theorem in $D$ dimensional Quantum Field Theory.}
In this section we review  CPT transformations and the CPT theorem
in even-dimensional 
Minkowski space-time in the context of quantum field theory.

The CPT theorem [\Ref{Pauli},\Ref{Luders},\Ref{SW},\Ref{others}] 
asserts that any 
quantum field theory is 
invariant under CPT transformations assuming that it satisfies the 
assumptions a), b) and c) stated in the Introduction.
These assumptions also imply the spin-statistics theorem, i.e. fields of 
integer (half odd integer) spin are 
quantized with respect to Bose (Fermi) statistics.

The CPT transformation actually comes in two varieties, one being  
the hermitean conjugate of the other: 

The first, which is what Pauli [\Ref{Pauli}] called 
{\it strong reflection} (SR), 
essentially maps a quantum field $\phi(t,\vec{x}) \rightarrow ({\rm 
phase}) \ \phi(-t,-\vec{x})$, i.e. it reverses time and space coordinates 
simultaneously. It also reverses the order of operators in an operator 
product and therefore cannot be represented by any 
operator acting on the particle states. 
It is a symmetry of the operator algebra only.

The second, which we will consider 
to be the CPT transformation proper, is obtained by performing first 
SR and then the hermitean conjugation (HC).
The resulting operation clearly does not change the order of 
operators in an operator product but instead all c-numbers are complex 
conjugated. It can be represented by an anti-unitary 
(i.e. unitary and anti-linear) operator $\Theta$ acting on the Hilbert 
space of physical particle states. 

As is well known, the combination of the transformations
of charge conjugation, parity and time reversal (C~$+$~P~$+$~T), when 
defined, is equal to the combination of strong reflection and 
hermitean conjugation (SR~$+$~HC),\note{This is true provided that the phase for
each transformation is chosen appropriately.} thereby justifying the name 
CPT for the latter transformation.
\section{CPT invariance in lagrangian field theory}
It is relatively easy to demonstrate CPT invariance in the framework of 
Lagrangian field theory if we assume the validity of the spin-statistics 
theorem. Our starting point is then given by the hypothesis of 

\itemitem{a)}{Lorentz invariance,}

\itemitem{d)}{spin-statistics theorem.}

\noindent
Following L\"{u}ders [\Ref{Luders}], we start by defining the SR and CPT 
transformations on field operators that obey 
free field equations of motion and free field (anti-) commutation relations.
In any (even) number D of space-time dimensions these are as follows:
\newline 
For a scalar field:
$$\eqalignno{ & (\partial_{\mu} \partial^{\mu} - m^2 ) \phi (x) = 0 & 
\nameali{scalar} \cr
& [ \phi^{\dagger} (x), \phi(x') ] = -i \Delta(x-x') \ . \cr }$$
For a spinor field:
$$\eqalignno{ 
& (\gamma^{\mu} \partial_{\mu} + m   ) \psi (x) = 0  & \nameali{spinor} \cr
& \{ \psi (x), \overline{\psi} (x') \} = i(\gamma^{\mu} \partial_{\mu} - m) 
\Delta (x-x') \ .  \cr }$$
For a vector field:
$$\eqalignno{   
& (\partial_{\mu} \partial^{\mu} - m^2 ) 
\phi^{\nu} (x) = 0  \qquad \qquad \partial_{\mu} 
\phi^{\mu} (x) = 0 & \nameali{vector} \cr
& [ \phi^{\dagger}_{\mu} (x), \phi_{\nu}(x') ] = -i (\eta_{\mu \nu} - {1 \over 
m^2} \partial_{\mu} \partial_{\nu}) \Delta(x-x') \ , \cr }$$
where our notation is to take $x=(t,\vec{x})$, the metric $\eta = 
{\rm diag}(-1,+1,\ldots,+1)$ and the gamma matrices to satisfy
$$\eqalignno{ & \{ \gamma^{\mu}, \gamma^{\nu} \} = 2 \eta^{\mu \nu} 
\qquad \qquad (\gamma^{\mu})^{\dagger} = \gamma_{\mu} & 
\nameali{gammaprop} \cr
& \overline{\psi} (x) = \psi^{\dagger}(x) (-i \gamma_0) \ = \ 
\psi^{\dagger}(x) (i \gamma^0) \cr 
& C (\gamma^{\mu})^T C^{-1} = - \gamma^{\mu} \qquad \qquad
C^{\dagger} = C^{-1} \ . \cr   } $$
The Schwinger function $\Delta$ is given by
$$ \Delta(x) = i \int {d^{D-1} \vec{p} \over (2\pi)^{D-1} 2 p^{0}} 
\left( e^{i p \cdot x} - e^{-i p \cdot x} \right) \quad {\rm 
with}\ \ \ p^0 = \sqrt{\vec{p}^2 + m^2} \nfr{Schwinger}
and clearly satisfies
$$\Delta(-x) = -\Delta(x) \ . \efr
Under SR the individual operators transform as follows
$$\eqalignno{ & \phi(x) \ \srarrow \ \phi(-x) & \nameali{sr} \cr
& \psi(x) \ \srarrow \ \varphi_{_{\rm SR}} \gamma^{D+1} \psi(-x) \qquad 
{\rm with} 
\qquad (\varphi_{_{\rm SR}})^2 = -(-1)^{D/2} \cr
& \overline{\psi}(x) \ \srarrow \ - \varphi_{_{\rm SR}}^* \overline{\psi}(-x) 
\gamma^{D+1} \cr
& \phi^{\mu} (x) \ \srarrow \ - \phi^{\mu} (-x) \ , \cr } $$
where $\gamma^{D+1}$ is the chirality matrix (normalized up to a sign by 
the requirement that the square is the unit matrix).
It is easy to verify that the equations of motion and (anti-) commutation 
relations \scalar\ -- \vector\ are invariant under these transformations 
when we also define 
SR to {\it invert the order of the operators} and {\it 
leave c-numbers unchanged}. At this point it is obviously
essential that the fields satisfy the spin-statistics
relation.

For complex fields the requirement that the (anti-) commutation relations
are invariant under SR only fixes the form of the SR transformations 
\sr\ up to an overall phase factor; but for real fields the choice of 
$\varphi_{_{\rm SR}}$ is constrained by the requirement that the 
SR transformation is consistent with hermitean conjugation and
only a sign ambiguity remains.~\note{Real (Majorana) fermions of nonzero 
mass can only be introduced in 
dimensions $D=2+8k$ or $D=4+8k$, $k \in {\Bbb N}$.} 
The transformation laws given in 
\sr\ correspond to a choice of phase that is consistent also for real 
fields. The dependence on the dimension that enters into the phase 
$\varphi_{_{\rm SR}}$ for the spinor field is due to the fact that the charge 
conjugation matrix commutes with $\gamma^{D+1}$ in dimensions
$D= 4k, k \in {\Bbb N}$ but anti-commutes in dimensions 
$D=2+4k, k\in {\Bbb N}$.

For the bosonic fields the sign is chosen to agree with what we 
obtain by applying the tensor transformation law to the transformation
$x \rightarrow -x$. Thus a field with $N$ vector indices would transform under 
SR with a phase $(-1)^N$. 

By definition, the CPT transformation is obtained as the SR 
transformation \sr\ followed by the operation of hermitean conjugation. 
Unlike the SR transformation, CPT may be represented by an anti-unitary 
operator $\Theta$ acting on the free (multi-)particle states. Each free
single-particle state is completely characterized by  three labels 
$$\vert\rho\rangle =\vert p, \eta, \{\lambda\}\rangle \ , 
\nfr{statebasis}
where $p$ is the momentum of the state, $\eta$ describes the spin 
degrees of freedom and 
$\{\lambda\}$ is a collective label
for all charges and enumerative indices carried by the state. 

In $D=4$, we may identify $\eta$ with the helicity, defined as the 
projection of the spin along the momentum; more precisely as the 
eigenvalue of the operator 
$H={\vec{J}\cdot\vec{p}\over\vert\vec{p}\vert}$, see also Appendix 
A.~\note{In higher (even) dimensions, $\eta$ may be thought of as the 
eigenvalues of the Lorentz generators $M^{12}, M^{34}, \ldots , 
M^{D-3,D-2}$, if the momentum points in the $(D-1)$-direction. We always 
use the word ``helicity'' for short.} 

The CPT transformation 
flips the helicity and the charges, leaving the momentum invariant. 
More precisely
$$ \vert \rho \rangle \ = \ \vert p, \eta, \{ \lambda \} \rangle \ 
\cptarrow \ \vert \rho^{\rm CPT} \rangle \ = \ \varphi_{_{\rm CPT}} (\eta, 
\{ \lambda \} ) \ \vert p, -\eta, \{  -\lambda \} \rangle \ , \nfr{physSR}
where the phase $\varphi_{_{\rm CPT}} (\eta ,\{\lambda \})$ 
may depend on the quantum numbers $\eta$ and $\{ \lambda \}$, as 
indicated. Multiparticle states transform like a direct product of 
single-particle states:
$$ \vert \rho_{_{1}};\rho_{_{2}};\ldots;\rho_{_{N}} \rangle \ \cptarrow \ \vert 
\rho_{_{1}}^{\rm CPT};\rho_{_{2}}^{\rm CPT};\ldots;\rho_{_{N}}^{\rm CPT} 
\rangle \ ,
\nfr{physSRmulti}
and
$$ \left( \vert \rho_{_{1}};\rho_{_{2}};\ldots;\rho_{_{N}} \rangle 
\right)^{\dagger} \ = \ \langle \rho_{_{N}};\ldots;\rho_{_{2}};\rho_{_{1}} 
\vert \ . \efr
Finally, the phase of the vacuum is chosen so that 
$$ \vert 0 \rangle \ \cptarrow \ \vert 0 \rangle \ .
\nfr{vacuumSR}
Having defined the SR and CPT transformations for free quantum fields 
(and hence for free multiparticle states) 
we turn our attention to an interacting theory. 
In this case the quantum fields are no longer free, but they can still 
formally be expressed in terms of the free fields by means of the 
time-evolution operator $U(t)$.

To show that SR remains a symmetry even in the interacting case, 
we must show that the interaction Lagrangian ${\cal L}_{\rm int} (x)$
is invariant under the free-field transformations \sr . 
To this end we note that any 
object carrying $N$ Lorentz vector indices transforms under SR with a 
sign $(-1)^N$. For the fundamental bosonic quantum fields this holds by 
definition; for the derivative $\partial_{\mu}$ it is trivial, and by 
inspection it holds as well for any (pseudo-)tensor of the form
$:\overline{\psi} \gamma^{\mu_1 \ldots \mu_n} \psi:$ or
$:\overline{\psi} \gamma^{\mu_1 \ldots \mu_n} \gamma^{D+1} \psi:$. Thus,
any Lorentz-invariant 
interaction Lagrangian that we can build by taking normal-ordered 
products of the fields and their derivatives will be invariant 
under SR.
 
Under CPT, the interaction Lagrangian and the $S$-matrix
then transform according to
$$ \Theta {\cal L}_{\rm int} (x) \Theta^{-1} \ = \ \left( {\cal L}_{\rm 
int} (-x) \right)^{\dagger} \quad {\rm and} \quad 
\Theta S \Theta^{-1} \ = \ S^{\dagger}
\ . \nfr{CPTLagrange}
Since $\Theta$ is an anti-linear operation we can consider \CPTLagrange\ to 
express CPT invariance of the interacting theory regardless of whether 
the interaction Lagrangian is hermitean or not. Thus, SR invariance is 
considered equivalent to CPT invariance. Of course, unless the 
interaction Lagrangian is hermitean, the $S$-matrix  
will not be unitary and the concept of asymptotic states will be 
rather meaningless. For this reason we will assume unitarity to be 
satisfied in what follows, keeping in mind that it is strictly speaking 
a property distinct from that of SR/CPT invariance.

For a general vacuum expectation value of local field operators, 
the statement of SR/CPT invariance becomes
$$
\eqalignno{ \langle 0 \vert\Phi_1 (x_1) & \ldots\Phi_n (x_n)\vert 0
\rangle  & \nameali{cptinvariance} \cr
& \eqcpt \ \langle 0 \vert  
\Theta \Phi_1 (x_1) \Theta^{-1} \ldots 
\Theta \Phi_n (x_n) \Theta^{-1}  \vert 0 \rangle^*  \cr
& \ = \ \langle 0 \vert \left( \Phi_n (x_n) \right)^{\rm SR} 
\ldots \left( \Phi_1 (x_1) \right)^{\rm SR} \vert 0 
\rangle \cr
& \ = \ (-1)^{J_1} (-1)^{J_2} N \langle 0 \vert \Phi_n (-x_n) 
\ldots \Phi_1 (-x_1) \vert 0 \rangle \ , \qquad\qquad \cr } 
$$
where $(\Phi_i (x_i))^{\rm SR}$ is defined by the free-field 
transformation laws \sr , 
$J_1$ is the number of Lorentz vector indices, $J_2$ is the number 
of $\gamma^{D+1} = -1$ spinor indices, and
$$ 
N = \left\{ \matrix{ i^{N_F} & {\rm if} & D = 4k, k \in 
{\Bbb N} \cr 
& & \cr  
(-1)^{N_{\downarrow}} & {\rm if} & D = 2+4k, k \in 
{\Bbb N} \cr}   \right. \ ,  
\efr
with $N_F$ being the total number of fermions (which is even) and 
$N_{\downarrow}$ the number of covariant spinor indices (our 
convention is that $\psi$ is contravariant and $\overline{\psi}$ is 
covariant). The different behavior in $2$ and $4$ dimensions (mod $4$) 
is again due to the different chirality properties of the charge 
conjugation matrix. 

If the points $x_1,\ldots,x_n$ are such that all separations $x_i-x_k$, 
$i \neq k$, are space-like, and if we assume the validity of the 
spin-statistics theorem, the following condition, sometimes called {\it 
Weak Local Commutativity}, holds~[\Ref{SW}]:
$$ 
\langle 0 \vert \Phi_n (x_n) \ldots \Phi_1 (x_1) \vert 0 \rangle =
i^{N_F} \langle 0 \vert \Phi_1 (x_1) \ldots \Phi_n (x_n) \vert 0 
\rangle \ . \nfr{wlc}
The phase appearing is just a sign, counting how many times two fermions 
have been transposed in the process of reversing the order of the 
operators. A priori this sign is $(-1)^{N_F(N_F-1)/2}$, but since $N_F$ is  
even this equals $i^{N_F}$.

Combining eqs. \cptinvariance\ and \wlc\ we obtain another formulation 
of CPT invariance:
$$ \eqalignno{
& \langle 0 \vert \Phi_1 (x_1) \ldots \Phi_n (x_n) \vert 0 \rangle 
\ = & \nameali{cptandwlc} \cr &\qquad\qquad 
i^{N_F} \langle 0 \vert \left( \Phi_1(x_1) \right)^{\rm SR} \ldots 
\left( \Phi_n (x_n) \right)^{\rm SR} \vert 0 \rangle \ , \cr} $$ 
valid when all separations are space-like.
\section{CPT invariance of the $S$-matrix elements.}
Anticipating the situation in string theory (where we have no 
fully-fledged Lagrangian formulation) we proceed to consider 
directly the $S$-matrix elements. 
Under CPT the asymptotic states $\vert \rho_{_{1}}; \ldots ; \rho_{_{N}} 
; {}^{in}_{out} \rangle$ transform in the same way as the free 
multiparticle states, i.e. according to eq.~\physSRmulti , with an 
additional interchange of the ``{\it in}'' and ``{\it out}'' labels.
In terms of the $S$-matrix elements the statement of 
field theory CPT invariance becomes
$$\eqalignno{
\langle \rho_{_{1}}; 
\ldots; & \rho_{_{N_{out}}};in  \vert S\vert\rho_{_{N_{out}+1}};
\ldots;\rho_{_{N}};in \rangle &\nameali{Scpt}\cr
& \eqcpt \ \langle \rho_{_{1}}^{\rm CPT}; \ldots; \rho_{_{N_{out}}}^{\rm CPT} 
;out \vert S^{\dagger} \vert\rho_{_{N_{out}+1}}^{\rm CPT};
\ldots;\rho_{_{N}}^{\rm CPT}; out \rangle^*  \cr
&= \ \langle\, \rho_{_{N}}^{CPT}; 
\ldots; \rho_{_{N_{out}+1}}^{CPT};in 
\vert S\vert\, \rho_{_{N_{out}}}^{CPT} ;\ldots;  \rho_{_{1}}^{CPT};in\rangle 
\ . \cr}
$$ 
To mimic the situation in string theory as closely as possible we 
represent the $S$-matrix elements
by means of the Lehmann-Symanzik-Zimmermann reduction formula~[\Ref{IZ}]
as in ref.~[\Ref{normaliz}]:
$$
\eqalignno{& \langle \rho_{_{1}}, \ldots , \rho_{_{N_{out}}}; 
in \vert S \vert \rho_{_{N_{out}+1}}, \ldots, 
\rho_{_N} ;in 
\rangle\ = \ {\rm disconnected \ terms}\ +  &\nameali{LSZvo}\cr
&\qquad \left(\prod^{N}_{j=1} {i \over \sqrt{Z_j}} 
\right)  \int \left(\prod_{j=1}^{N}
{\di}^4{x}_j \right)   \langle 0 \vert {\rm T} 
V_{\langle \rho_1 \vert} (x_1)  \ldots
V_{\vert \rho_{N} \rangle} (x_{_{N}}) 
\vert 0 \rangle \ , \cr}
$$
where we have a Field Theory Vertex (FTV) 
$V_{\vert \rho \rangle} (x)$ corresponding 
to the $1$-particle ket-state $\vert \rho ; in \rangle$ 
and similarly a FTV $V_{\langle \rho \vert} (x) = (V_{\vert \rho \rangle} 
(x))^{\dagger}$ corresponding to the 1-particle bra-state 
$\langle \rho ; in \vert$. 

Using the identity \cptinvariance\ on the correlation function appearing 
in \LSZvo\ one finds that the statement of CPT (or SR) invariance can be 
formulated as follows~\note{Since we will mainly consider states of the 
``$in$'' variety, unless otherwise stated 
$\vert \rho\rangle = \vert \rho;in\rangle$, i.e. we drop the ``$in$'' label.}
$$\eqalignno{ &\langle\, \rho_{_{1}}; \ldots; \rho_{_{N_{out}}} 
\vert S\vert\rho_{_{N_{out}+1}};
\ldots;\rho_{_{N}} \rangle 
\ = &\nameali{expone} \cr 
&\qquad{\rm disconnected \ terms}\ +\ 
\left(\prod^N_{j=1} {i \over \sqrt{Z_j}} \right)\  
\int \left(\prod_{j=1}^N {\di}^4{x}_j \right)\ \times\cr   
&\qquad \langle 0 \vert {\rm T} 
\left( V_{\vert\rho_N \rangle}(x_{_{N}})
\right)^{SR}\ldots \left(V_{\langle \rho_1 \vert} (x_1)\right)^{SR}
\vert 0 \rangle \ . \cr }
$$
On the other hand, we may also use eq.~\LSZvo\ to rewrite
$$\eqalignno{ & \langle \, \rho_{_{N}}^{\rm CPT}; \ldots; 
\rho_{_{N_{out}+1}}^{\rm CPT} 
\vert S \vert \rho_{_{N_{out}}}^{\rm CPT};
\ldots ; \rho_{_{1}}^{\rm CPT} \rangle 
\ = & \nameali{exptwo} \cr 
&\qquad{\rm disconnected \ terms}\ +\ 
\left(\prod^{N}_{j=1} {i \over \sqrt{Z_j}} \right)\  
\int \left(\prod_{j=1}^{N}{\di}^4{x}_j \right)\ \times\cr   
& \qquad \langle 0 \vert {\rm T} 
V_{\langle \rho_N^{\rm CPT} \vert}(x_{_{N}})
\ldots V_{\vert \rho_1^{\rm CPT} \rangle} (x_1) 
\vert 0 \rangle \ . \cr }$$
Since the $S$-matrix element appearing on the left-hand side of 
eq.~\expone\ equals the one appearing on the left-hand side of 
eq.~\exptwo\ (by the identity \Scpt ) it follows that the right-hand 
sides of these two equations must coincide as well. Since this is true 
for any combination of states, it follows that
$$ \left( V_{\vert \rho \rangle} (x) \right)^{\rm SR} = V_{\langle 
\rho^{\rm CPT} \vert} (-x) = \left( V_{\vert \rho^{\rm CPT} \rangle} 
(-x) \right)^{\dagger} \ , \nfr{cptft} 
which can be viewed as an indirect way of defining the CPT transformed 
one-particle state $\vert \rho^{\rm CPT} \rangle$ directly from the 
known transformation properties under SR
of the FTV $V_{\vert \rho \rangle}$. This point of view will turn out to 
be very useful for our formulation of the CPT transformation in string 
theory.
\section{The CPT theorem for two-dimensional conformal field theory.}
For the purpose of string theory, we are interested in world-sheets with 
a two-dimensional metric, a compactified space-direction and possibly a 
non-trivial topology. Since this is incompatible with 2-dimensional 
Lorentz invariance, the CPT theorem in its standard form, as described 
in subsection 1.1, is not directly applicable.

However, we are only interested in 2-dimensional {\it conformal} field 
theories, where the role of the SR transformation is naturally taken by 
the Belavin-Polyakov-Zamolodchikov (BPZ) transformation~[\Ref{BPZ}]
$$ z \rightarrow w = {1 \over z} \ , \nfr{bpztrans}
which defines a globally conformal diffeomorfism on the sphere. 
In terms of the cylindrical coordinates $(\tau,\sigma)$ we have 
$z=\exp\{\tau + i \sigma\}$ and the BPZ transformation \bpztrans\ is 
seen to map $(\tau,\sigma) \rightarrow (-\tau,-\sigma)$, as desired.
A primary conformal field $\Phi_{(\Delta,\overline{\Delta})}$ of conformal
dimension $(\Delta,\overline{\Delta})$ transforms 
as
$$\eqalignno{ \Phi_{(\Delta,\overline{\Delta})} (z=\zeta,\bar{z}=\bar{\zeta}) 
\ \bpzarrow & \ \Phi_{(\Delta,\overline{\Delta})} 
(w=\zeta,\bar{w}=\bar{\zeta}) \ & \nameali{bpz} \cr
& = \ (-1)^{\Delta-\overline{\Delta}}
\left( {1 \over \zeta^2} \right)^{\Delta} 
\left( {1 \over \bar{\zeta}^2} \right)^{\overline{\Delta}} 
\Phi_{(\Delta,\overline{\Delta})} (z=1/\zeta,\bar{z}=1/\bar{\zeta}) \ ,
\cr}$$
which involves a choice of phase whenever $\Delta - \overline{\Delta}$ 
is not an integer. 
Conformal invariance implies the identity~\note{For notational 
convenience we restrict ourselves to 
chiral conformal fields in the rest of this subsection.}
$$ \langle \ \Phi_{\Delta_1} (z=\zeta_1) \ldots \Phi_{\Delta_n} 
(z=\zeta_n) \ \rangle \ = \ \langle \ \Phi_{\Delta_1} (w=\zeta_1) \ldots
\Phi_{\Delta_n} (w=\zeta_n) \ \rangle \ . \nfr{bpz}
On the sphere it is also possible to introduce the operation of 
hermitean conjugation, for example at the level of the mode operators. 
We define
$$ \left( \Phi_{\Delta} (z=\zeta) \right)^{\dagger} \ = \
\left( {1 \over \zeta^*} \right)^{2\Delta} \widehat{\Phi}_{\Delta} 
(z=1/\zeta^*) \ , \nfr{hat}
where $\widehat{\Phi}_{\Delta}$ is a primary conformal field of the same 
dimension as $\Phi_{\Delta}$. The field $\Phi_{\Delta}$ is said to be 
hermitean if $\Phi_{\Delta} = \widehat{\Phi}_{\Delta}$ and 
anti-hermitean if $\Phi_{\Delta} = -\widehat{\Phi}_{\Delta}$.
The peculiar behaviour of the argument in eq.~\hat\ 
is due to the fact that we are considering imaginary time on the 
world-sheet. Rotating back to real time, we have $z=\zeta=\exp\{ i(\tau 
+\sigma)\}$ and $1/\zeta^* = \zeta$.

By means of the hermitean conjugate, eq.~\bpz\ can be reformulated as 
follows
$$\eqalignno{ & \langle \ \Phi_{\Delta_1} (z=\zeta_1) \ldots \Phi_{\Delta_n} 
(z=\zeta_n) \ \rangle \ = \ \langle \ \left( \Phi_{\Delta_n} (w=\zeta_n) 
\right)^{\dagger} \ldots \left( \Phi_{\Delta_1} (w=\zeta_1) 
\right)^{\dagger} \ \rangle^*  \cr
& \qquad = \ (-1)^{\Delta_1} \ldots (-1)^{\Delta_n} \ \langle \
\widehat{\Phi}_{\Delta_n} (z=\zeta_n^*) \ldots \widehat{\Phi}_{\Delta_1} 
(z=\zeta_1^*) \ \rangle^* \ , & \nameali{twodcpt} \cr} $$
and unlike the individual transformations of BPZ and hermitean 
conjugation, which are defined on the sphere, the identity between the 
first and the last correlator in eq.~\twodcpt\ 
actually holds at {\it any} genus for a {\it unitary} conformal field 
theory~[\Ref{Sonoda},\Ref{MooreSeiberg},\Ref{uswcpt}] 
and is referred to as CPT 
invariance in two dimensions. It expresses the symmetry between a given 
Riemann surface and its anti-holomorphic ``mirror image''. It differs 
from the standard formulation \cptinvariance\ of CPT invariance in 
having the opposite ordering of the operators appearing in the complex 
conjugated correlator. As such it is more akin to the relation 
\cptandwlc\ expressing the combination of
CPT invariance and Weak Local Commutativity.

Along similar lines it is tempting to define a two-dimensional 
(world-sheet) CPT 
transformation by
$$ \Phi_{\Delta} (z=\zeta) \ \wcptarrow \ \left( \Phi_{\Delta} (w=\zeta) 
\right)^{\dagger} \ = \ (-1)^{-\Delta} \widehat{\Phi}_{\Delta} 
(z=\zeta^*) \nfr{twodcpttrans}
but one should keep in mind that 
this is only a substitution rule, rather than a genuine CPT 
transformation, inasmuch as it involves an 
inversion of the ordering of operators and therefore cannot be 
represented by any operator acting on states. 

For non-chiral conformal fields the phase in eq.~\twodcpttrans\ is 
replaced by $(-1)^{\overline{\Delta}-\Delta}$.

\chapter{The space-time CPT theorem for heterotic strings 
in a Minkowski background}
In this section we consider first-quantized heterotic
string models in a $D$-dimensional Minkowski background ($D$ even),
having a unitary $S$-matrix and satisfying Lorentz-invariance as well as
the space-time spin-statistics theorem in a form that we will formulate 
more precisely below. As we pointed out already in the 
Introduction, and saw in more detail in section 1, 
these assumptions suffice to prove the CPT 
theorem in quantum field theory. We will show that this is also the case 
in string theory. Our procedure is to first define an SR-like 
transformation on the fields defined on the world-sheet, which is a 
symmetry of the underlying 2-dimensional conformal field theory; 
next, we construct the space-time CPT transformation on the set of 
physical string states by means of the SR symmetry and finally we prove 
that with this definition of CPT conjugate 
string states, the identity \Scpt\ holds for the string $S$-matrix 
elements. 

To describe the heterotic superstring we use the Neveu-Schwarz Ramond 
(NSR) formalism but it will become obvious that our results
actually hold more generally, for example also for string models
in the Green-Schwarz formalism. In the NSR formalism we have 
various free conformal fields: The space-time coordinates $X^{\mu}$, 
their chiral world-sheet superpartners $\psi^{\mu}$, the 
reparametrization ghosts $b,c$ and $\bar{b},\bar{c}$, and the 
superghosts $\beta,\gamma$. On top of this we have various internal 
degrees of freedom described by an ``internal'' CFT with 
left-moving (right-moving) central charge 22 (9). These may or may not 
be free. 
\section{The space-time SR transformation}
We define the space-time 
SR transformation in the string CFT in the obvious way, 
taking
$$ X^\mu \srarrow -\, X^\mu\ , \qquad\qquad \psi^\mu \srarrow -\, \psi^\mu 
\ , \nfr{srXpsiSone}
with the (super) ghosts, as well as all fields pertaining to the 
``internal'' CFT, remaining unchanged. The SR transformation 
\srXpsiSone\ is {\it 
linear} (i.e. it does not complex conjugate $c$-numbers); unlike its 
quantum field theory namesake {\it it does not invert the order of 
operators}. This convention is seen to be the sensible one, since only
thus defined will the SR transformation leave 
invariant the world-sheet action of the string theory, the BRST current, 
and the world-sheet fermion number operators obtained from the currents 
$\psi^0 \psi^1$ and $i\psi^2 \psi^3$, and hence the GSO 
projection conditions.

Next we define how SR acts on the vacuum (or vacua) of any given sector of the 
string theory. There are two kinds of sectors. Those containing states 
of integer space-time spin, where the vacuum transforms as a Lorentz 
singlet, and those containing states of half odd integer space-time 
spin, where it transforms as a Lorentz spinor. The vacuum 
may also be parametrized by various other quantities $\{ \lambda \}$, 
pertaining to the ``internal'' CFT; from the space-time point of view 
these quantities are interpreted as charges and various labels. We 
define
$$\eqalignno{ 
& \vert 0 ; \{ \lambda \} \rangle \srarrow \vert 0 ; \{ \lambda \} 
\rangle \qquad \hbox{\rm for a sector of integer spin,} & 
\nameali{SRvacuum} \cr 
& \vert \alpha ; \{ \lambda \} \rangle \srarrow \varphi_{_{\rm SR}} 
\ ( \gamma^{D+1} )_{\alpha}^{\  \beta} \ \vert \beta ; \{ \lambda \} 
\rangle \qquad \hbox{\rm for a sector of half odd integer spin,} \cr } 
$$
where $\varphi_{_{\rm SR}}$ is some as yet unspecified phase factor.
The presence of the $\gamma^{D+1}$ is needed in order to obtain an 
unambiguous result for the SR transform of the state
$$ \psi^{\mu}_0 \ \vert \alpha ; \{ \lambda \} \rangle = {1 \over 
\sqrt{2}} (\gamma^{\mu})_{\alpha}^{\ \beta} \ \vert \beta ; \{ \lambda 
\} \rangle \ , \efr
where $\psi_0^{\mu}$ is the zero-mode of the field $\psi^{\mu}$.
We may introduce spin fields that create the various vacua from the 
conformal one; 
collecting all indices of the vacuum into a single one, 
${\Bbb A}$, we may write
$$ S_{\Bbb A} (z=\bar{z}=0) 
\vert 0 \rangle \equiv \vert {\Bbb A} \rangle \ . \efr
In terms of the spin fields the transformation laws \SRvacuum\ become 
$$\eqalignno{ 
& S_{\Bbb A} \srarrow S_{\Bbb A} 
\qquad \hbox{\rm for a sector of integer spin,} & 
\nameali{SRspinfield} \cr 
& S_{\Bbb A} \srarrow \varphi_{_{\rm SR}} \left( {\bfmath \Gamma}^{D+1} 
\right)_{\Bbb A}^{\ \, {\Bbb B}} \ S_{\Bbb B} \qquad \hbox{\rm for a 
sector of half odd integer spin,} \cr }  
$$
where obviously, writing ${\Bbb A} = (\alpha; \{ \lambda \} )$ and
${\Bbb B} = (\beta; \{ \lambda' \} )$, we have defined
$$ \left( {\bfmath \Gamma}^{D+1} 
\right)_{\Bbb A}^{\ \, {\Bbb B}} = (\gamma^{D+1})_{\alpha}^{\ \beta} \
\delta_{\{ \lambda \} }^{\ \{ \lambda' \} } \ . \efr
The phase factor $\varphi_{_{\rm SR}}$ is fixed up to a sign by the 
requirement that SR should commute with hermitean conjugation.
Indeed, if we bosonize the fermions $\psi^{\mu}$, introducing
$D/2$ scalar fields $\phi_{(0)}, \phi_{(1)}, \ldots , \phi_{(D/2-1)}$, 
the spin field operator $S_{\Bbb A}$ pertaining to a sector of half odd 
integer spin may be written as
$$ S_{\Bbb A} = e^{a_0 \phi_{(0)}} e^{a_1 \phi_{(1)}}  \ldots e^{a_{D/2-1} 
\phi_{(D/2-1)}} \times ( {\rm internal \ part} ) \times ( {\rm cocycle \ 
factor } ) \ , \efr
where all the quantities $a_0,a_1, \ldots $ take values $\pm 1/2$ and
the space-time spinor index $\alpha = (a_0,a_1,\ldots,a_{D/2-1})$.  
In this formulation the matrix $\gamma^{D+1}$ may be thought of as a 
direct product of $D/2$ matrices $\sigma_3^{(i)}$, which take 
values $+1$ ($-1$) depending on whether the number $a_i=+1/2$ ($-1/2$).

When the bosonization is performed in Minkowski space-time, the operator 
field $\phi_{(0)}$ is hermitean, whereas the other ones, $\phi_{(1)}, 
\ldots, \phi_{(D/2-1)}$, are anti-hermitean~[\Ref{mink}]. 
Therefore, when we take the 
hermitean conjugate of the spin field, we obtain a new spin field, where
the sign has changed on all the $a_i$ except $a_0$. Accordingly, the 
value of $\gamma^{D+1}$ has changed by the sign $(-1)^{D/2-1}$.

Therefore, the requirement that the operations of SR and hermitean 
conjugation should commute leads to the constraint
$$ \varphi_{_{\rm SR}}^* = \varphi_{_{\rm SR}} (-1)^{D/2-1} 
\nfr{SRphaseone}
or equivalently
$$ (\varphi_{_{\rm SR }})^2 = - (-1)^{D/2} \ , \nfr{SRphase}
which agrees with the phase choice found in field theory when one 
requires compatibility of SR with a Majorana reality condition (q.v. eq.~\sr). 
 
In the following we will always assume the phase $\varphi_{_{\rm SR}}$ to 
be given in accordance with \SRphase.

The SR transformation is a symmetry of the conformal 
field theory describing the space-time fields $X^{\mu}$, since it 
leaves both the action and the path-integral measure invariant. 
Therefore we have the identity
$$ \langle ({\cal O}_1)^{\rm SR} (z_1,\bar{z}_1) \ldots ({\cal 
O}_N)^{\rm SR} (z_N,\bar{z}_N) 
\rangle = \langle {\cal O}_1 (z_1,\bar{z}_1) \ldots {\cal 
O}_N (z_N,\bar{z}_N) 
\rangle \ , \nfr{XSRinvariance}
which holds for any correlation function in the CFT of the fields $X^{\mu}$.

If we proceed to consider the CFT of all the other fields in
the string theory, including the fields $\psi^{\mu}$, together 
with the ``internal'' CFT and the (super) ghosts, 
the SR transformation is again clearly a 
symmetry of the action. However, when the fields $\psi^{\mu}$ contain
branch cuts (corresponding to the presence of pairs of spin fields 
transforming as space-time spinors), 
the path integral measure is no longer SR invariant. 
Instead we obtain the modified SR identity
$$ \langle ({\cal O}_1)^{\rm SR} (z_1,\bar{z}_1) \ldots ({\cal 
O}_N)^{\rm SR} (z_N,\bar{z}_N) 
\rangle = (-1)^{N_{\rm FP}}\ \langle {\cal O}_1 (z_1,\bar{z}_1) \ldots {\cal 
O}_N (z_N,\bar{z}_N) 
\rangle \ , \nfr{SRinvariance}
where $2N_{\rm FP}$ is the number of space-time spinorial spin 
fields.~\note{That the number of space-time spinorial spin-fields is
even for nonzero correlators is a consequence of the assumption of 
space-time Lorentz invariance. No invariant tensor exists with an odd 
number of spinor indices.}

To prove this, we do not have to specify in detail how the path-integral 
measure is defined. 
It is sufficient to assume that it is defined to 
be invariant under Lorentz transformations continuously connected to the 
identity.

Indeed, consider --in the operator formulation-- the correlation 
function involving $2 N_{\rm FP}$ fields
$$ \left( {\cal O}_i^F \right)^{\mu_1^{(i)} \dots 
\mu^{(i)}_{n_i}}_{\alpha_i} \qquad i = 1, \ldots, 2N_{\rm FP} 
\nfr{opone}
that carry a space-time spinor index $\alpha_i$, and $N_B$ operator 
fields
$$ \left( {\cal O}_i^B \right)^{\nu_1^{(i)} \dots 
\nu^{(i)}_{m_i}} \qquad i = 1, \ldots, N_{\rm B} \nfr{optwo}
that do not. The fields may also carry various vector indices, as shown. 
They will in general also carry all sorts of indices connected with the 
``internal'' CFT, but these are all suppressed as they do not play any 
role in the present discussion. We imagine all Lorentz indices in 
\opone\ and \optwo\ to be 
carried by the world-sheet operators; no Dirac 
spinors, polarization or momentum vectors etc. have been introduced.
Being c-numbers, they do not transform under SR anyway.

The assumption of invariance of 
the path integral under infinitesimal Lorentz transformations implies 
first, that the correlation function of the operators \opone\ and \optwo\
transforms under such transformations in 
accordance with the index structure of the operators, i.e. as an object 
with $2 N_{\rm FP}$ covariant spinor indices and $n_1 + \ldots + 
n_{_{2N_{\rm FP}}} + m_1 + \ldots + m_{_{N_B}} \equiv N_V$ contravariant 
vector indices, and second, that the correlation function can be
expressed only by means of the invariant tensors at our disposal
$$ \eta_{\mu \nu}, \, \eta^{\mu \nu}, \, \epsilon^{\mu_1 \mu_2 \ldots \mu_D},
\, \left( \gamma^{\mu} \right)_{\alpha}^{\ \beta}, \, C_{\alpha \beta} \ 
, \nfr{tensors}
where $C_{\alpha \beta}$ is the spinor metric. We do not need to 
include $\gamma^{D+1}$ explicitly in the list, since $\gamma^{D+1} = 
{1 \over D!}
\varphi_{_{\rm SR}} \epsilon_{\mu_1 \mu_2 \ldots \mu_D} \gamma^{\mu_1} 
\gamma^{\mu_2} \ldots \gamma^{\mu_D}$.~\note{Since 
$(\varphi_{_{\rm SR}})^2 
=(-1)^{D/2-1}$, up to a sign this is the usual definition of
$\gamma^{D+1}$ in (even) $D$-dimensional Minkowski space-time.}
Also notice that for $D$ even 
there exist two spinor metrics, but one is obtained from the other 
merely by multiplying with $\gamma^{D+1}$. We may remove this ambiguity 
by taking $C$ to be the charge conjugation matrix, i.e. satisfying
$$ C (\gamma^{\mu})^T C^{-1} = - \gamma^{\mu} \ . \nfr{ccm}
Finally notice that we do not 
need to include the inverse spinor metric, because the only non-trivial 
way this could possibly enter is with both indices contracted with gamma 
matrices, so that $C^{-1}$ is multiplied from the right by a gamma 
matrix and from the left by a transposed gamma matrix. But using 
eq.~\ccm\
we may always move any such factor of $C^{-1}$ to the left, until we 
reach the end of the string of gamma matrices; and since all upper 
spinor indices are contracted, it will then inevitably cancel against a 
factor of $C$.

Having made these observations it is clear that 
to get the right number of lower spinor indices our correlator must 
contain exactly $N_{\rm FP}$ factors of the spinor metric. 
Each spinor metric may be multiplied by gamma matrices from the left and 
by transposed gamma matrices from the right. By repeated use of 
eq.~\ccm\ we may bring $C$ to the right, ``un-transposing'' all 
gamma-matrices in the process. This way we end up with 
$N_{\rm FP}$ structures of the type
$$ \left( \gamma^{\mu_1} \ldots \gamma^{\mu_n} C \right)_{\alpha \beta} 
\ .
\nfr{cstructure}
Now, under SR each of the operators \opone\ picks up a factor 
$(-1)^{n_i} \varphi_{_{\rm SR}} \gamma^{D+1}$ (acting on the spinor index
$\alpha_i$), while the operators \optwo\ all pick up the factor 
$(-1)^{m_i}$. Thus, the $N_V$ vector indices give rise to the sign 
$(-1)^{N_V}$. Since
$$\eqalignno{  (\varphi_{_{\rm SR}})^2 \gamma^{D+1} \gamma^{\mu_1} \ldots 
\gamma^{\mu_n} C \gamma^{D+1} \ &= \ 
(\varphi_{_{\rm SR}})^2 (-1)^n (-1)^{D/2} \gamma^{\mu_1} \ldots 
\gamma^{\mu_n} C\qquad & \numali \cr
& = \ - (-1)^n \gamma^{\mu_1} \ldots \gamma^{\mu_n} C \ , \cr } 
$$
the $N_{\rm FP}$ structures of the type \cstructure\ give rise to
a further sign which is just $(-1)^{N_{\rm FP}}$, times 
$(-1)^{N_G'}$, where $N_G'$ is the number of gamma matrices that appear in 
the expression for the correlator inside structures of the type 
\cstructure . Lorentz invariance only allows gamma matrices to appear 
without an accompanying spinor metric in one way, namely as traces 
$(\gamma^{\mu_1} \ldots 
\gamma^{\mu_n})_{\alpha}^{\ \alpha}$, but these are only nonzero if 
$n$ is even, so we may write $(-1)^{N_G'} = (-1)^{N_G}$, where $N_G$ is 
the total number of gamma matrices appearing. Since the metric and the 
Levi-Civita tensor always carry an even number of Lorentz vector 
indices, we may in fact write $(-1)^{N_G'} = (-1)^{N_G} = (-1)^{N_V^{\rm 
tot}}$, where $N_V^{\rm tot}$ is the total number of Lorentz vector 
indices appearing in the expression for the correlator, counting both 
covariant and contravariant indices. Some of these indices will in 
general be summed over. It is clear that $N_V^{\rm tot} = N_V + N_D$, 
where $N_D$ is the number of dummy indices, which is even, since any 
dummy index appears exactly twice. Therefore 
$(-1)^{N_V} = (-1)^{N_G}= (-1)^{N_G'}$, and we find that 
SR transforms the correlator 
into itself times a phase factor that finally reduces to 
$(-1)^{N_{\rm FP}}$. 

This concludes our proof of eq.~\SRinvariance .
A priori this identity holds for the CFT {\it excluding} the fields 
$X^{\mu}$, but since the CFT of the latter satisfies the identity 
\XSRinvariance , it is clear that eq.~\SRinvariance\ actually
holds for the entire CFT underlying the string theory in question.

In summary, the SR transformation we have defined 
can be considered to act on the space-time 
indices carried by the world-sheet fields and the identity \SRinvariance\ 
follows from the one basic assumption that the string models we consider 
are invariant under space-time Lorentz 
transformations continuously connected to the identity. In the following 
two subsections we show how the SR transformation can be used to define a CPT 
transformation on the space of physical string states and to prove the 
invariance of the $S$-matrix under this transformation.
\section{The space-time CPT transformation}
Having introduced the space-time SR transformation we now turn our 
attention to the definition of the space-time CPT transformation.
But first we need to introduce some
very general notation for the string $S$-matrix elements.

We define the $T$-matrix element (loosely referred to as ``the amplitude'')
as the connected $S$-matrix element with certain normalization factors removed
$$
\eqalignno{
& { \langle \rho_{_{1}}, \dots , \rho_{_{N_{\rm out}}} \vert S \vert
\rho_{_{N_{\rm out}+1}}, \dots , \rho_{_{N}} 
\rangle_{\rm connected} \over \prod_{i=1}^{N} 
\left( \langle \rho_i \vert \rho_i \rangle \right)^{1/2} }  
= & \nameali{Smatrix} \cr
&\qquad\quad i (2\pi)^D \delta^D (p_1 + \dots p_{_{N_{\rm out}}} - 
p_{_{N_{\rm out}+1}} - \dots - p_{_{N}}) 
\prod_{i=1}^{N} (2 p^0_i V)^{-1/2}\ \times \cr
&\qquad\quad T(\rho_{_{1}}; \dots ; \rho_{_{N_{\rm out}}} \vert 
\rho_{_{N_{\rm out}+1}}; \dots ; \rho_{_{N}} 
) \ , \cr } $$
where $p_i$ is the momentum of the $i$'th string state, all 
of them having $p_i^0 > 0$, and $V$ is the 
usual volume-of-the-world factor.  We also introduce the dimensionless 
momentum $k_\mu \equiv \sqrt{{\alpha^\prime \over 2}} \, p_\mu$.
The Minkowski metric is ${\rm diag}(-1,1,\ldots ,1)$. 

Corresponding to each on-shell 
single-string state $\vert \rho \rangle$ appearing 
in the $T$-matrix element we have a BRST-invariant vertex
operator ${\cal W}_{\vert \rho \rangle} (z,\bar{z})$, defined
by
$$ \vert \rho \rangle \ = \ \lim_{\zeta,\bar{\zeta} \rightarrow 0}
{\cal W}_{\vert \rho \rangle} (z=\zeta,\bar{z}=\bar{\zeta}) 
\ \vert 0 \rangle \ , \efr
which is a primary conformal field of dimension 
$\Delta=\overline{\Delta}=0$.

Similarly, as discussed in more details in ref. [\Ref{normaliz}], 
to each state $\langle\rho\vert$ we associate a vertex operator 
${\cal W}_{\langle \rho \vert} (z,\bar{z})$, which --apart 
from a phase factor $\chi$, depending on the superghost 
charge (i.e. the picture) of the state $\vert \rho \rangle$--
is simply obtained from the vertex operator ${\cal W}_{\vert \rho \rangle}$ 
by means of the
world-sheet CPT transformation \twodcpttrans : 
$$\eqalignno{ {\cal W}_{\langle \rho \vert} (z=\zeta,\bar{z}=\bar{\zeta}) 
&  = \  \chi \ 
\left( {\cal W}_{\vert \rho \rangle} (z=\zeta^*,\bar{z}=\bar{\zeta}^*) 
\right)^{\rm WS-CPT} & \nameali{inoutmap} \cr
& = \ \chi \
\widehat{{\cal W}}_{\vert \rho \rangle} (z=\zeta,\bar{z}=\bar{\zeta}) \ 
. \cr}$$
Whereas the operator ${\cal W}_{\vert \rho \rangle}$ 
creates a ket state of positive 
energy, i.e. is proportional to $e^{i k \cdot X}$ with $k^0 > 0$, the 
operator ${\cal W}_{\langle \rho \vert}$ is 
proportional to $e^{-i k \cdot X}$ and thus creates a ket state with 
negative energy. 

As it is well known, the $T$-matrix element is given by the formula
$$ 
\eqalignno{ & T (\rho_{_{1}}; \dots ; \rho_{_{N_{\rm out}}} \vert
\rho_{_{N_{\rm out}+1}};\dots;\rho_{_{N}} ) \ = \ \sum_{g=0}^{\infty} 
C_g \ \int {\rm d} \mu \ \sum_{_{{}^{\rm \ \ \ spin}_{\rm structures} }}
\ \ \ & \nameali{Tmatrix} \cr
&\qquad\quad\qquad\wew{
 \left( {}^{\rm ghost}_{\rm insertions}  \right) \ 
\left( {}^{\rm picture \ changing}_{\rm \ \ \ \ operators}  \right) \ 
{\cal W}_{\langle \rho_1 \vert} 
(z_1,\bar{z}_1) \dots {\cal W}_{\vert \rho_{_N} \rangle}(z_{_N},\bar{z}_{_N}) 
}
\ ,\cr} 
$$
which involves a formal sum over topologies (plus, in the NSR
formulation of superstring theory, a summation over spin structures), 
and an integral over moduli, where the integrand is given by the 
correlator of the vertex operators, with various ghost and picture 
changing operators inserted. For what follows we do not need to specify 
the details in the formula \Tmatrix, which anyway depend on which kind 
of string theory we consider. The exact form taken by this formula in 
the specific instance of heterotic string theory can be found for example
in ref.~[\Ref{normaliz}].

We are now ready to define the space-time CPT transformation in 
string theory. 
At the level of the vertex operators we define, in close 
analogy with eq.~\cptft ,
$$ \left( {\cal W}_{\vert \rho \rangle} \right)^{\rm SR} \equiv
{\cal W}_{\langle \rho^{\rm CPT} \vert} \ . \nfr{CPTdef}
This definition makes sense: The vertex operator ${\cal W}_{\vert \rho 
\rangle}$ has positive energy; the SR transformed operator therefore has 
negative energy ($\exp\{ i k \cdot X\} \srarrow \exp\{ -i k \cdot X\}$) 
and may be thought of as 
the vertex operator pertaining 
to an outgoing string state $\langle \rho^{\rm CPT} \vert$. To find the 
state $\vert \rho^{\rm CPT} \rangle$ or, equivalently, the operator 
${\cal W}_{\vert \rho^{\rm CPT} \rangle}$, we simply use the map
\inoutmap\ ``backwards'', to obtain
$$ {\cal W}_{\vert \rho \rangle} \ \cptarrow \
{\cal W}_{\vert \rho^{\rm CPT} \rangle} = \chi \ \widehat{{\cal 
W}}_{\langle \rho^{\rm CPT} \vert} = \chi \left( \widehat{{\cal W}}_{\vert 
\rho \rangle} \right)^{\rm SR} = \left( {\cal W}_{\langle \rho \vert} 
\right)^{\rm SR}\ .
\nfr{CPTtrans}
Thus, the {\it space-time} CPT transformation mapping ${\cal W}_{\vert 
\rho \rangle}$ into ${\cal W}_{\vert \rho^{\rm CPT} \rangle}$ is 
essentially the combination of the SR transformation defined by 
eqs.~\srXpsiSone\ and \SRspinfield\ with the {\it world-sheet} CPT 
transformation \twodcpttrans\ which relates ${\cal W}_{\vert \rho \rangle}$ 
to ${\cal W}_{\langle \rho \vert}$, as in eq.~\inoutmap .
Notice that since hermitean conjugation and SR have been arranged to 
commute (by the phase choice \SRphase ) no ambiguity arises in the 
definition \CPTtrans\ by interchanging the order of the SR and world-sheet CPT 
transformations.

The transformation 
$\vert \rho \rangle \cptarrow \vert \rho^{\rm CPT} 
\rangle$  that we have defined is clearly anti-linear, due to the 
presence of hermitean conjugation in the definition \hat\ of the ``hatted'' 
operator.
As was shown in ref.~[\Ref{normaliz}]
the operator ${\cal W}_{\langle \rho \vert}$ satisfies the requirement 
of BRST invariance and the GSO conditions if and only if ${\cal 
W}_{\vert \rho \rangle}$ does. Since the SR 
transformation defined in the previous subsection leaves the BRST current 
and the GSO conditions invariant, the conclusion is that the state
$\vert \rho^{\rm CPT} \rangle$, defined by eq.~\CPTtrans , is BRST 
invariant and in the GSO projected spectrum if and only if $\vert 
\rho \rangle$ is.
\section{Proof of the space-time CPT theorem}
We are now ready to prove the 
space-time CPT theorem in string theory. 
The ingredients will be Lorentz invariance, as encoded in 
the identity \SRinvariance , together with 
{\it space-time spin-statistics}, as expressed by the following basic 
assumption: 

For any pair of
physical, BRST-invariant vertex operators ${\cal W}_{\vert \rho_1 \rangle}$
and ${\cal W}_{\vert \rho_2 \rangle}$, describing incoming string 
states in the GSO projected spectrum that can appear together in a 
nonzero amplitude, it is required that
$$ {\cal W}_{\vert \rho_1 \rangle} (z_1,\bar{z}_1)
{\cal W}_{\vert \rho_2 \rangle} (z_2,\bar{z}_2) \ = \
\pm {\cal W}_{\vert \rho_2 \rangle} (z_2,\bar{z}_2)
{\cal W}_{\vert \rho_1 \rangle} (z_1,\bar{z}_1) \ , \nfr{spinstatws}
the sign being minus if both $\vert \rho_1 \rangle$ and $\vert \rho_2 
\rangle$ carry half odd integer space-time spin and plus otherwise.

By the relation \inoutmap\ it is 
clear that also the operators ${\cal W}_{\langle \rho_1 \vert}$ and
${\cal W}_{\langle \rho_2 \vert}$, or ${\cal W}_{\vert \rho_1 \rangle}$ 
and ${\cal W}_{\langle \rho_2 \vert}$, will then
satisfy Bose or Fermi statistics on the world-sheet
depending on whether their space-time spin
is integer or half odd integer. 

At the level of the amplitudes given by eq.~\Tmatrix\ our assumption 
\spinstatws\ ensures that
$$ \eqalignno{ & T( \rho_{_{1}}; \ldots ; \rho_{_{N_{\rm out}}} \vert
\rho_{_{N_{\rm out}+1}}; \ldots ; \rho_i; \rho_{i+1}; \ldots ; 
\rho_{_{N}} )\ = & \nameali{spinstat} \cr
& \qquad\qquad\qquad \pm T( \rho_{_{1}}; \ldots ; \rho_{_{N_{\rm out}}} \vert
\rho_{_{N_{\rm out}+1}}; \ldots ; \rho_{i+1}; \rho_{i}; \ldots ; 
\rho_{_{N}} ) \cr } $$
(and similarly for interchange of outgoing string states), with the sign 
being minus if the single-string states $\vert \rho_i \rangle$ and 
$\vert \rho_{i+1} \rangle$ both describe string states with half odd 
integer space-time spin, and plus otherwise.

It is now straightforward to prove the space-time CPT invariance \Scpt\ 
of the string $S$-matrix elements at any loop order: 
$$\eqalignno{ T( \rho_{_{N}}^{\rm CPT}; \ldots ; \rho_{_{N_{\rm 
out}+1}}^{\rm CPT} \vert &\rho_{_{N_{\rm out}}}^{\rm CPT}; \ldots ; 
\rho_1^{\rm CPT} ) \ = & \nameali{CPTproof}\cr 
&=\sum \int \wew{ (\ldots) {\cal W}_{\langle \rho_{_{N}}^{\rm CPT} \vert } 
\ldots {\cal W}_{\vert \rho_1^{\rm CPT} \rangle} } \cr
&=\  \sum \int \wew{ (\ldots)^{\rm SR} 
\left({\cal W}_{\vert \rho_{_{N}} \rangle 
}\right)^{\rm SR} \ldots \left({\cal W}_{\langle \rho_1 \vert} 
\right)^{\rm SR} } \cr
&=\  (-1)^{N_{\rm FP}} \sum\int\wew{(\ldots) {\cal W}_{\vert\rho_{_N}\rangle 
} \ldots {\cal W}_{\langle\rho_1\vert}  }  \cr
&=\ (-1)^{N_{\rm FP}} (-1)^{N_{\rm FP} (2N_{\rm FP}-1)} \
\sum \int \wew{ (\ldots) {\cal W}_{\langle \rho_1 \vert} \ldots
{\cal W}_{\vert\rho_{_N}\rangle} }  \cr
&=\ T(\rho_1; \ldots ; \rho_{_{N_{\rm out}}} \vert \rho_{_{N_{\rm 
out}+1}}; \ldots ; \rho_{_N} )\ . \cr } $$
In going from the second to the third line we used the definition 
\CPTdef\ of the CPT transformation (and the inverse relation \CPTtrans 
), together with the fact that the various ghost and picture changing 
operators $(\ldots)$ are SR invariant; 
Next we used the SR identity \SRinvariance , which 
followed from the assumption of Lorentz invariance; Finally, we used the
spin-statistics assumption \spinstatws\ to
invert the order of the vertex operators. Inverting the order of 
the $2N_{\rm FP}$ vertex operators of half odd integer spin
produces the sign $(-1)^{N_{\rm FP}(2N_{\rm FP}-1)}$ 
which cancels the sign appearing in the SR identity \SRinvariance.

This concludes the proof of the space-time CPT theorem. 

\section{Comments on the proof of the space-time CPT theorem}
As we have already noticed, the SR transformation, 
defined in string theory by eqs. \srXpsiSone, \SRspinfield\ and \SRphase , 
is sensitive only to the Lorentz spacetime 
indices carried by the world-sheet operators. As such it does not depend 
on the details of the explicit string model under consideration and is 
of very general application. 

Also the assumptions used in the proof of the spacetime CPT theorem are
very general. We considered a $D$-dimensional ($D$ even) string theory 
on a Minkowski background and we required 
\item{$\bullet$}{unitarity of the $S$-matrix,}
\item{$\bullet$}{explicit Lorentz invariance of the string theory,}
\item{$\bullet$}{validity of the space-time spin-statistics theorem 
\spinstatws\ for 
the physical string state spectrum.}

\noindent
The assumption of unitarity of the $S$-matrix does not play any direct
role in the proof of the space-time CPT theorem as we have formulated it.
Indeed, as we pointed out in section 1, the question of space-time
CPT invariance is strictly speaking independent of the assumption of unitarity.
But without unitarity, the $S$-matrix would
of course make little physical sense and 
most of our statements would be valid only at the formal level. 

We also note that, at the level of the string amplitudes, 
CPT invariance and unitarity appear in rather different ways. 
As we have seen, CPT invariance 
is implemented at the level of the vertex operator correlation function, 
i.e. is an invariance of the modular integrand, q.v. eq.~\CPTproof. By 
contrast, unitarity only appears {\it after} integrating over the 
moduli, with appropriate regularizations of short-distance 
divergencies~[\Ref{Hoker},\Ref{Berera},\Ref{hokgid},\Ref{Weisberger}].

To illustrate this point further, we may consider the pseudo eletric dipole 
moment (PEDM) encountered in ref. [\Ref{ammedm}]. This was interpreted 
as a potential CPT-violating term in the amplitude under investigation. 
Actually, a careful analysis shows that, in a 
Minkowski background, the PEDM, if nonzero, would be imaginary and thus
violate unitarity rather than CPT. Indeed, it was found to be zero only 
as a result of integrating over the moduli.

Finally we would like to stress that, since in the proof of the 
spacetime CPT theorem we did not need to 
specify in much detail how to evaluate the scattering amplitudes, our 
result can even be considered to be {\it non-perturbative} as far as 
one is able to give a non-perturbative interpretation of the string 
scattering amplitudes 
\Tmatrix\ which preserves all the assumptions of the theorem.
\chapter{The physical interpretation of the space-time CPT transformation
in string theory.}
In the previous section we defined the CPT transformation by 
eqs.~\CPTdef\ and \CPTtrans , and checked that it leads to the 
invariance \CPTproof\ of the string amplitudes. 
But it still remains to be checked that 
the would-be CPT transformation has the correct space-time 
interpretation. 

Indeed, taking the field theory limit 
($\alpha^\prime \rightarrow 0$), any string
theory gives rise to a field theory whose spectrum is given by the massless
spectrum of the string theory. This requires that the CPT transformation
in string theory goes into the field theory one for the string massless
states. 

We already know that the 
state $\vert \rho^{\rm CPT} \rangle$ is in the physical spectrum if and 
only if $\vert \rho \rangle$ is. 
We want to show that for any given single-string state 
$\vert \rho \rangle = \vert k,\eta,\{\lambda\}\rangle$, 
created by a vertex operator
${\cal W}_{\vert \rho \rangle}$, having helicity $\eta$ and 
charges/labels $\{ \lambda \}$, the CPT transformed state $\vert 
\rho^{\rm CPT} \rangle$ has helicity $-\eta$ and charges/labels $\{ - 
\lambda \}$, i.e. is proportional to
$\vert k,-\eta,\{-\lambda\}\rangle$. For mere labels the change of sign 
is to a certain extent a matter of convention, but for charges 
pertaining to a gauge symmetry group it is a necessity.

For notational simplicity we will restrict ourselves to the case $D=4$
in what follows, but our arguments are readily generalized to other 
(even) values of $D$.

We start with a general discussion which we will then clarify with
two explicit examples given in the following subsections.

In string theory any gauge charge $\lambda$ (i.e. any 
weight pertaining to some generator of the Cartan 
subalgebra) is an eigenvalue of the zero mode operator 
$\Lambda $ obtained from the corresponding hermitean Ka\v{c}-Moody 
current, $J_{\Lambda}$, that involves only the ``internal'' CFT.
That the string state has charge $\lambda$ is equivalent to saying that 
the first order pole in the operator product expansion of $J_{\Lambda}$ 
with ${\cal W}_{\vert \rho \rangle}$ is as follows:
$$ J_{\Lambda} (z=\zeta_1) {\cal W}_{\vert \rho \rangle} 
(z=\zeta_2,\bar{z}=\bar{\zeta}_2) \eqtope {\lambda \over \zeta_1 - 
\zeta_2} {\cal W}_{\vert \rho \rangle} (z=\zeta_2,\bar{z}=\bar{\zeta}_2)  
\ . \nfr{kacmoodyope}
We saw in the previous section that the space-time CPT transformation is 
essentially composed of the SR transformation and the world-sheet CPT 
transformation.

As is seen from the definition \twodcpttrans , the world-sheet CPT 
transformation changes sign on any chiral hermitean field of dimension 
one, in particular on any Ka\v{c}-Moody current, and so it follows, by 
performing the world-sheet CPT transformation on both sides of 
eq.~\kacmoodyope , that if the operator ${\cal W}_{\vert \rho \rangle}$ 
carries charge $\lambda$, then the operator ${\cal W}_{\langle \rho 
\vert} = \chi \widehat{\cal W}_{\vert \rho \rangle}$ carries charge 
$-\lambda$.

Since the SR transformation defined by eqs.~\srXpsiSone , \SRspinfield\ 
and \SRphase \ only affects the space-time degrees of freedom, it is 
clear that if the operator ${\cal W}_{\langle \rho \vert}$ carries charge 
$-\lambda$, so does the operator 
$({\cal W}_{\langle \rho \vert})^{\rm SR} = 
{\cal W}_{\vert \rho^{\rm CPT} \rangle}$, and hence the state $\vert 
\rho^{\rm CPT} \rangle$.

Thus we see that the sign change on $\lambda$ is a result, not of the SR 
transformation, but of the world-sheet CPT transformation, i.e. of the 
map \inoutmap\ relating the vertex operator ${\cal W}_{\vert \rho 
\rangle}$ and ${\cal W}_{\langle \rho \vert}$.

As regards the helicity, we define this to be the inner product of the 
angular momentum with a unit vector pointing in the same direction as 
the eigenvector of the four-momentum operator, as explained in greater 
detail in Appendix A. 
Under the world-sheet CPT transformation relating
${\cal W}_{\vert \rho \rangle}$ to 
${\cal W}_{\langle \rho \vert}$, both the momentum and the angular momentum 
in the direction of the momentum
(being eigenvalues pertaining to the Ka\v{c}-Moody currents of translations and 
rotations respectively) are flipped, 
meaning that the helicity is unaffected.
Under SR the four-momentum operator flips sign but the 
angular momentum does not (since the position four-vector is also 
flipped); accordingly the operator $({\cal W}_{\langle \rho \vert})^{\rm 
SR} = {\cal W}_{\vert \rho^{\rm CPT} \rangle}$ creates a state with 
helicity $-\eta$.  In 
conclusion, the state $\vert \rho^{\rm CPT} \rangle$ has the opposite 
helicity than $\vert \rho \rangle$, as desired.

This concludes our proof that the map $\vert \rho \rangle \rightarrow 
\vert \rho^{\rm CPT} \rangle$ does indeed have the space-time properties 
of a CPT transformation. Notice that as such 
it automatically defines for us the 
concept of {\it anti-particle} in string theory. Simply, the CPT 
transformed of a string state 
$\vert\rho\rangle = \vert k,\eta,\{ \lambda \} \rangle$ is (up to a phase) 
equal to the ``anti-particle'' string state of opposite helicity, 
$\vert \rho^{\rm CPT} \rangle \propto \vert k,-\eta,\{ -\lambda \} \rangle$.  
Of course, the anti-particle state with 
the {\it same} helicity, i.e. $\vert k, +\eta, \{ -\lambda \} \rangle$, 
needs not be in the physical spectrum. This depends on whether both signs
of the helicity are allowed by the GSO projection conditions.

By construction the states $\vert \rho \rangle$ and $\vert \rho^{\rm 
CPT} \rangle$ have the same (bare) mass. Space-time CPT invariance 
guarantees that this equality holds also for the renormalized 
masses.

It is worthwhile to stress here the difference between world-sheet and 
space-time CPT. As was pointed out by Witten [\Ref{Wittwcpt}] the 
world-sheet CPT transformation relates a string state with given 
charge and mass to a string state with opposite charge and equal mass.
So does the space-time CPT transformation we have defined. But whereas 
the world-sheet CPT transformation also changes sign on the energy of 
the string state, the space-time CPT transformation does not. Therefore, 
the space-time interpretation of the two transformations is very 
different: World-sheet CPT maps the vertex operator pertaining to an 
incoming string state into the vertex operator pertaining to the {\it 
same} string state but outgoing. 
Instead space-time CPT maps the vertex operator 
of an incoming ``particle'' string state into the vertex operator of an 
incoming ``anti-particle'' string state (of opposite helicity).
\section{The gluon vertex operator.}
In the first of our two examples we consider a ``gluon'', described in a $D=4$ 
dimensional heterotic string theory
by the vertex operator~[\Ref{normaliz},\Ref{ammedm}]
$$ {\cal W}_{\vert \rho \rangle} = {\kappa \over \pi} c \bar{c} \bar{J}^a 
\epsilon \cdot \psi e^{-\phi} e^{i k \cdot X} \ , 
\nfr{gluon}
where $\kappa$ is the gravitational coupling ($\kappa^2 = 8 \pi G_N$ in 
$D=4$), $\bar{J}^a$ is the hermitean Kac-Moody current, normalized so 
that $\bar{J}^a (\bar{z}) \bar{J}^b (\bar{w}) = \delta^{ab} (\bar{z} - 
\bar{w})^{-2} + \ldots$, and $e^{-\phi} = \delta(\gamma)$. For this
picture the phase $\chi=+1$ in eq. \inoutmap\ [\Ref{normaliz}], and
$$\eqalignno{ {\cal W}_{\vert \rho \rangle} \cptarrow& 
{\cal W}_{\vert\rho^{\rm CPT}\rangle}\ =\ \left({\cal W}_{\langle\rho\vert}
\right)^{SR}\ =\ \left(\widehat{\cal W}_{\vert\rho\rangle}\right)^{SR}\ =
& \nameali{CPTgluon} \cr
=&\left( {\kappa \over \pi} e^{-i k \cdot X} e^{-\phi} \epsilon^* 
\cdot \psi \bar{J}^a\bar{c} c \right)^{\rm SR}\cr 
=& \left( {\kappa \over \pi} c \bar{c}  \bar{J}^a  \epsilon^* \cdot \psi 
e^{-\phi} e^{-i k \cdot X}   \right)^{\rm SR} \cr
=& - {\kappa \over \pi} c \bar{c} \bar{J}^a \epsilon^* \cdot \psi 
e^{-\phi} e^{i k \cdot X} \ . \cr}
$$
If we assume $\epsilon$ to describe a photon of definite helicity
(say, $\eta=+1$), 
then $\epsilon^*$ describes a photon of helicity $\eta=-1$ and thus the 
CPT transformed state is indeed proportional to the photon state with 
unchanged momentum but opposite helicity. 

If instead we assume the photon to have linear polarization in a given 
direction, $\epsilon^{\mu} = \delta^{\mu}_{\nu}$, then CPT maps the state 
into minus itself.
This is of course the correct behaviour well known from quantum field 
theory~[\Ref{IZ}]. 
\section{Ground states of world-sheet free fermion (KLT) models.}
We now consider four-dimensional heterotic string models of the 
Kawai-Lewellen-Tye (KLT) type [\Ref{KLT},\Ref{Anto}], where the internal 
degrees of freedom are described by 22 left-moving and 9 right-moving 
free complex fermions, which are labelled by $L=1,\ldots,22;23,\ldots,31$. 
In any given sector of the string theory a 
general state of the conformal field theory
(excluding the reparametrization ghosts) can be built by means of 
non-zero mode creation operators acting on the states $\vert k, {\Bbb A} 
\rangle$ which are obtained by adding a nonzero
space-time momentum $k$ to the vacuum state(s) $\vert {\Bbb A} \rangle$ 
introduced in subsection 2.1.

If we bosonize all the 31 ``internal'' complex fermions (as well as the 
four Majorana fermions $\psi^{\mu}$), using the explicit prescription 
for bosonization in Minkowski space-time proposed in ref.~[\Ref{mink}], 
the collective label ${\Bbb A}$ simply consists of
the vacuum values of the ``momenta'' $J_0^{(L)} = {\Bbb A}_L$ pertaining 
to the 33 bosons $\Phi_{(L)}$ introduced by the bosonization, 
and the vacuum superghost charge
$J_0^{(34)}=q={\Bbb A}_{34}$ which is (minus) the ``momentum'' 
of the field $\phi \equiv \Phi_{(34)}$ that is introduced when ``bosonizing'' 
the superghosts. With the standard choice of picture, $q=-1$ for 
sectors of integer space-time spin
and $q=-1/2$ for sectors of half odd integer space-time spin. 

Since $[ J_0^{(L)} , \Phi_{(K)} ] = \delta^{L}_{\ K}$, 
the operator creating the state $\vert k, {\Bbb A} \rangle$
from the conformal vacuum is
$$ S_{\Bbb A} (z,\bar{z}) \ e^{i k \cdot X(z,\bar{z})} \ , 
\nfr{groundst}
where 
$$ S_{\Bbb A} (z,\bar{z}) \equiv \prod_{L=1}^{34} e^{{\Bbb A}_L 
\Phi_{(L)} (z,\bar{z})} \left( C_{(L)} \right)^{{\Bbb A}_L}  
\nfr{spinfield}
is a spin field operator and $C_{(L)}$ is a cocycle factor, see 
ref.~[\Ref{mink}] for details.

The values of the ${\Bbb A}_L$ in the vacuum (i.e. the values that minimize 
$L_0 + \bar{L}_0$) depend on the sector and hence on the details 
of the KLT model we happen to consider, see 
refs.~[\Ref{KLT},\Ref{ammedm}]. Actually we may as well
consider a slightly more general set of states, where the ``internal'' 
momenta ${\Bbb A}_L$, $L=1,\ldots,31$, assume any of the values allowed 
in the given sector, not just the values corresponding to the vacuum.
But for the two ``momenta'' ${\Bbb A}_{32}$ and ${\Bbb A}_{33}$ 
pertaining to the bosonization of the $\psi^{\mu}$ 
we still restrict ourselves to the vacuum proper, 
meaning that in sectors of integer spin we 
have ${\Bbb A}_{32}={\Bbb A}_{33}=0$, so that the state $\vert {\Bbb A} 
\rangle$ is a space-time scalar,  whereas in 
sectors of half odd integer spin
we have ${\Bbb A}_{32}=\pm 1/2$ and ${\Bbb A}_{33}=\pm 1/2$, 
so that the state $\vert {\Bbb A} \rangle$ transforms as a space-time 
spinor, with the four-dimensional
spinor index $\alpha$ is given by $\alpha=({\Bbb A}_{32},{\Bbb A}_{33})$.

We are interested in physical external states of this type, so we 
assume the level-matching condition 
$L_0 - \bar{L}_0 = 0$ to be satisfied and consider vertex operators 
given by
$$ \eqalignno{ &
{\cal W}_{\vert k,\eta,\{\lambda\} \rangle} (z,\bar{z}) \ = \
W_{(in)}^{\Bbb A} (k,\eta,\{\lambda\}) \, S_{\Bbb A} (z,\bar{z}) 
\ e^{i k \cdot X(z,\bar{z})} c(z) \bar{c} (\bar{z}) \ . &
\nameali{groundvert}\cr} 
$$
Here the $c$-number quantities $W^{\Bbb A}_{(in)} (k,\eta,\lambda)$ are 
partially fixed by the GSO conditions.
If the vertex operator in eq. \groundvert\ describes a
state of spin $1/2$ ($\eta=\pm 1/2$), 
$W_{(in)}^{\Bbb A} (k,\eta,\{\lambda\})$ is a space-time spinor and 
has superghost charge $-1/2$, i.e. is proportional to 
$\delta_{{\Bbb A}_{34},-1/2}$. It satisfies a Dirac equation which can 
be obtained by enforcing BRST invariance, more precisely, the 
requirement that the $3/2$-order pole in the 
operator product expansion (OPE) of the orbital part of the
supercurrent, $T_F^{[X,\psi]}$, with the operator \groundvert\ should vanish.
If we define the gamma matrices by the OPE
$$ \psi^{\mu} (z) S_{\Bbb A} (w,\bar{w}) \eqope {1 \over \sqrt{2}} 
\left( {\bfmath \Gamma}^{\mu} \right)_{\Bbb A}^{\ \Bbb B} S_{\Bbb B} 
(w,\bar{w}) {1 \over \sqrt{z-w}} + \ldots \ , \nfr{gammadef}
the Dirac equation assumes the matrix form
$$ (W_{(in)} (k,\eta,\{\lambda\}))^T \ {\Bbb D} \, (k) = 0 
\qquad {\rm or} \qquad
\left( {\Bbb D}\, (k) \right)^T W_{(in)} (k,\eta,\{\lambda\}) = 0 \ , 
\nfr{Diraceq}
where the Dirac operator is
$$ {\Bbb D}\, (k) = k_{\mu} {\bfmath \Gamma}^{\mu} - {\Bbb M} \ , \efr
${\Bbb M}$ being a mass operator that we do not need to write down 
explicitly.

Analogously, when the vertex operator 
in eq. \groundvert\ describes a space-time scalar ($\eta=0$), 
BRST invariance is 
reduced to the requirement that the second order pole in the 
operator product expansion (OPE) of $T_F^{[X,\psi]}$ 
with the operator \groundvert\ vanishes. Depending on the details of the 
``internal'' momenta this may be automatically satisfied; otherwise, it 
leads to an ``internal'' transversality condition on the $W^{\Bbb 
A}_{(in)} (k,\eta,\{\lambda\})$. 

Consider first the case of a spin $1/2$ string state. Under the space-time CPT
transformation given by eq.~\CPTtrans , the vertex operator \groundvert\ 
transforms as follows
$$
{\cal W}_{\vert k,\eta,\{\lambda\} \rangle} (z,\bar{z}) \cptarrow
\varphi_{_{SR}} (W_{(in)} (k,\eta,\{\lambda\}))^\dagger
{\bfmath \Gamma}^0 {\Bbb C}^{T,-1} {\bfmath \Gamma}_5
\, S (z,\bar{z}) \ e^{i k \cdot X(z,\bar{z})} c(z) \bar{c}(\bar{z}) \ , 
\nfr{spincpt}
where we used eq. (7.3) of ref. [\Ref{normaliz}] and defined
the charge conjugation matrix
$$ {\Bbb C}_{{\Bbb A} {\Bbb B}} =  \left(
\prod_{L=1}^{33} \delta_{{\Bbb A}_L + {\Bbb B}_L,0} \right) \
\delta_{{\Bbb A}_{34},{\Bbb B}_{34}} 
\ e^{ i \pi {\Bbb A} \cdot Y \cdot {\Bbb B}}\ , 
\nfr{Chargeconjq}
with $Y$ being the $34 \times 34$ cocycle matrix $Y_{KL}$ (see 
refs.~[\Ref{mink},\Ref{ammedm}]).

The state created by the CPT transformed vertex operator \spincpt\
has charges $\{-\lambda\}$ due to the presence of the charge conjugation
matrix, and, by using the explicit form of the helicity operator described
in Appendix A , has helicity $-\eta$. Thus it is
a state of the form $\vert k,-\eta,\{-\lambda\} \rangle$, 
as it was required for the correct
physical interpretation of the space-time CPT transformation. 

Analogous considerations can be done for space-time scalars and can be 
generalized to any other vertex operator in the theory. 
\chapter{The Spin-Statistics Relation}
One of the assumptions on which our proof of CPT invariance
relies is that for a physical string state,
the statistics of the vertex operator on the world-sheet should be 
given by the space-time spin of the string state, in accordance with the 
well-known spin-statistics theorem of quantum field theory 
[\Ref{otherSS},\Ref{SW},\Ref{DHR}].
More precisely, we required that two vertex operators (describing 
physical string states that can appear together in a nonzero amplitude) 
anti-commute if both carry half odd integer space-time spin and commute 
otherwise.

In field theory in more than two dimensions, 
the spin statistics theorem is well understood. It follows
directly from the basic axioms of field theory and the proof is quite 
parallel and related to the one of the CPT theorem itself [\Ref{SW}].
One way of looking at it is to say that the spin statistics theorem
is one of the steps required in the proof of the CPT theorem. 

In field theory the spin statistics theorem requires that a field of 
half odd integer spin is quantized according to canonical anti-commutation 
relations and a field of integer spin according to canonical commutation 
relations. An equivalent formulation is that the free-field contribution 
to the vacuum energy is positive (negative) for fields of integer (half 
odd integer) spin.
In its more general form the theorem does {\it not} 
require a given fermionic field to necessarily
anti-commute with {\it other} fermionic fields or a bosonic field to 
necessarily commute with all other fields, whether bosonic or fermionic. 
The situation where some field has abnormal statistics with respect to 
some of the other fields is referred to as {\it para-statistics\/}. 
However, one may show [\Ref{parastat}]
that in more than two dimensions
there always exists a map to a new basis of fields 
(a so-called Klein map [\Ref{Klein}]) 
which gives an equivalent physical system where all 
fields obey the standard spin-statistics relation [\Ref{SW}].

In string theory the situation is not so well understood. We
do not have an axiomatic framework (such as a fully-fledged string field 
theory) from which we could derive general 
theorems, this is why we had to prove the space-time CPT theorem
directly on the scattering amplitudes. So, if we do not want to simply
assume the spin-statistics relation and discard any string model which does
not satisfy it, the approach that we have taken up to now in this paper,
all what we can do is to discuss how the spin-statistics relation 
appears in string theory. We do not have major results to report and in
this section we will just try to illustrate a few points on this subject.

In building a Neveu-Schwarz Ramond-like string model, various
consistency requirements, in particular modular invariance, forces one 
to impose a sensible projection on the spectrum, the GSO projection [GSO]. 

In all known cases the GSO projections thus constructed also happen to
select the physical states in such a way that states of integer 
space-time spin contribute to the one-loop partition function with an 
overall plus sign, whereas states of half-odd integer space-time spin 
contribute with an overall minus sign~[seiberg-witten,Anto].  
Thus, all known consistent models require a GSO projection which implies 
a space-time spin-statistics relation.
As far as we know, a proof from first principles that this is the only
possibility, i.e. of the absence of pathological string theories violating the
spin-statistics relation, is not known.

Thus our expectation is that the GSO conditions should be crucial for 
obtaining the desired relation \spinstatws\ 
between space-time spin and world-sheet 
statistics of the physical state vertex operators. On the other hand, 
we do not expect the GSO conditions {\it alone} to guarantee such a relation.
One might imagine having para-statistics also in string theory, i.e. 
vertex operators pertaining to {\it different} string states might have 
unusual statistics with respect to one another. 

Indeed, we can always take the field theory limit of a string model 
thus obtaining a field theory whose spectrum is composed by the massless 
modes of the string spectrum. Since para-statistics is an a priori 
possibility in field theory, this argument, although it does not 
constitute a proof, strongly suggests the possibility of para-statistics 
also in string theory. 

So, we would like to understand if it is possible to build string models
satisfying para-statistics and, in this case, if there always exists a
physically equivalent string model satisfying the standard statistics. 
Should string models exist where this is not the case, 
the proof we have given of the CPT theorem would break down for such 
models.

We do not have a general answer to this question, but we can study what
happens in a particular class of string theories.

We restrict ourselves to the KLT models, that is, four dimensional
heterotic string models built with free world-sheet fermions 
[\Ref{KLT},\Ref{Anto},\Ref{Bluhm}]. As in
subsection 3.2 we bosonize all fermions. We want to compute the relative 
statistics of any two physical vertex operators in the model, i.e. find 
the phase that appears when the two operators are interchanged.
We may ignore the universal factor $c 
\bar{c} \exp\{ i k \cdot X \}$ which is always commuting. In fact, it 
is enough to consider string states created by the spin field operator 
$S_{\Bbb A}$, given by eq.~\spinfield , assuming that we let the ${\Bbb 
A}_L$, $L=1,\ldots,33$, take {\it any} of the values allowed in the 
string model (i.e. do not restrict ourselves to the vacuum values in 
each sector). This is because all
other states are obtained by means of 
the non-zero-mode creation operators of the fields $\Phi_{(L)}$, which 
are all commuting and do not affect either the spin (which is encoded in the 
values ${\Bbb A}_{32}$ and ${\Bbb A}_{33}$) or the GSO 
conditions, which only involve the fermion numbers, i.e. the values of
${\Bbb A}_L$, $L=1,\ldots,34$. As always ${\Bbb A}_{34} = -1~(-1/2)$ for 
sectors of integer (half odd integer) spin, and no superghost excitations 
above this vacuum are allowed for physical states.

By using the conventions of refs. [\Ref{mink},\Ref{normaliz},\Ref{ammedm}],
we can easily compute the phase one obtains when transposing two generic 
spin field operators in a KLT model. We get
$$
S_{{\Bbb A}_1} S_{{\Bbb A}_2} \ =\ e^{i\pi\sum_{K,L}{\Bbb A}_{1,L}
\widetilde{Y}_{L,K} {\Bbb A}_{2,K} }\, S_{{\Bbb A}_2} S_{{\Bbb A}_1} \ = 
\ e^{i \pi {\Bbb A}_1 \cdot \widetilde{Y} \cdot {\Bbb A}_2} \, S_{{\Bbb A}_2}
 S_{{\Bbb A}_1} \ , 
\efr
where the $34\times 34$ matrix $\widetilde{Y}$ has all elements equal to 
$+1$ or $-1$ and is obtained by anti-symmetrizing the lower-triangular
matrix $Y$, which describes the choice of
cocycles, and adding the diagonal, ${\rm diag} \widetilde{Y}$,
equal to $-\epsilon \ (=\pm1)$ for the first 22 entries as well as the 34th, 
and equal to $+\epsilon$ for the entries from the 23rd to the 33rd.
The matrix $\widetilde{Y}$ is subject to the constraints described in 
refs.~[\Ref{ammedm},\Ref{mink}] which ensure that the BRST current has 
well-defined statistics and that the Picture Changing operator, as well 
as all Ka\v{c}-Moody currents, have bosonic statistics w.r.t. any of the 
spin fields $S_{\Bbb A}$.

Thus everything depends on the phase
$$ \phi[{\Bbb A}_1, {\Bbb A}_2]\ \equiv\ \sum_{K,L}{\Bbb A}_{1,L}
\tilde{Y}_{L,K} {\Bbb A}_{2,K} \ = \ {\Bbb A}_1 \cdot \widetilde{Y} 
\cdot {\Bbb A}_2 \ , 
\nfr{statphase}
which a priori depends on the choice of cocycles as well as on
the two states we are considering.

First we want to verify that the operator $S_{\Bbb A}$
satisfies Bose or Fermi statistics with respect to itself, depending 
only on whether the space-time spin is integer or half-odd integer.
Thus we assume that ${\Bbb A}_1={\Bbb A}_2={\Bbb A}$.
Recalling that $\widetilde{Y}_{K,L}=-\widetilde{Y}_{L,K}$ for $K\neq L$ 
we can rewrite eq.~\statphase\ as
$$\eqalignno{
\phi[{\Bbb A},{\Bbb A}]\ &  =\ \sum_L {\Bbb A}_L \widetilde{Y}_{LL} {\Bbb 
A}_L \ = \ 2\epsilon \left( -{1 \over 2} \sum_{L=1}^{22} ({\Bbb A}_L)^2 
- {1 \over 2} ({\Bbb A}_{34})^2 + {1 \over 2} \sum_{L=1}^{33} ({\Bbb 
A}_L)^2 \right)  \cr
& = \ 2\epsilon\left(L_0[{\Bbb A}] -
\bar{L}_0[{\Bbb A}] - L_0^{(\beta\gamma)}[{\Bbb A}_{34}] -
\frac12({\Bbb A}_{34})^2 \right)\ . & \numali\cr} $$
Since all non-zero-mode creation operators in our bosonized formulation carry
integer values of $L_0$ and $\bar{L}_0$, level matching implies that
$ L_0[{\Bbb A}] - \bar{L}_0[{\Bbb A}]=$~integer, and since the phase
$\phi[{\Bbb A}_1,{\Bbb A}_2]$ is obviously defined modulus 2, the first two
terms cancel and we remain with
$$
\phi[{\Bbb A},{\Bbb A}]\eqmodtwo -2\epsilon\left( L_0^{(\beta\gamma)}
[{\Bbb A}_{34}] + \frac12({\Bbb A}_{34})^2 \right)\ .
\efr
Now it is enough to remember that the energy of the superghost system is
$L_0^{(\beta\gamma)}[{\Bbb A}_{34}] = -\frac12
{\Bbb A}_{34}({\Bbb A}_{34} + 2)$ to obtain
$$ \phi[{\Bbb A},{\Bbb A}]\eqmodtwo 2\epsilon\,{\Bbb A}_{34}
\eqmodtwo \left\{ \matrix{ 
0 \quad {\rm in \ a \ sector \ of \ integer \ spin \ \ \ \ \ \ 
\quad \quad } \cr
1 \quad {\rm in \ a \ sector \ of \ half \ odd \ integer \ spin} \cr } \right.
\ .
\efr
Thus level-matching alone leads to the correct spin-statistics
relation of the operator $S_{\Bbb A}$ with itself. 

Now we discuss the case of two different spin fields.
The analysis for a general four-dimensional KLT model is possible but 
technically very complicated, and since
our aim is merely to illustrate the issues involved by means of 
an example, we restrict ourselves to the simplest possible 
model in four dimensions, where all world-sheet fermions have the same spin 
structure, which can be either Ramond (R) or Neveu-Schwarz (NS).
In this simple model there is only one GSO condition, which can be 
written on the form
$$ {\bf V}_0 \cdot G \cdot {\Bbb A} \eqmodone \left\{ \matrix{
0 \qquad & {\rm in \ the \ NS \ sector} \cr
1/2 + k_{00} \ & {\rm  in \ the \ R \ sector} \cr} \right. \ . 
\nfr{GSOcond}
Here ${\bf V}_0$ is a 34-vector with all entries equal to $1/2$, $G$ is 
a diagonal 34 $\times$ 34 matrix given by
$$ G \ = \ {\rm diag}\, ((+1)^{22};(-1)^{11};+1) \ = \ -\epsilon \, 
{\rm diag} \widetilde{Y} \efr
and $k_{00}$ is a free parameter that equals either $0$ or $1/2$ (mod 
$1$).

 Obviously we
have three cases to consider: i) both vertex operators are in the NS sector
and describe states of integer spin, ii) both vertex operators are in the 
R sector and describe states of half odd integer spin, iii) one vertex 
operator is in the NS
sector and one in the R sector, describing then one state of integer and 
one of half odd integer spin. 

In case i) both ${\Bbb A}_{1,L}$ and ${\Bbb A}_{2,K}$ are integer so 
that
$$\eqalignno{ \phi[{\Bbb A}_1,{\Bbb A}_2] \  & = \ \sum_{K,L} {\Bbb 
A}_{1,L} \widetilde{Y}_{L,K} {\Bbb A}_{2,K} \ \eqmodtwo \ \sum_{K,L} {\Bbb 
A}_{1,L} {\Bbb A}_{2,K} & \numali \cr
& \eqmodtwo \ \left( \sum_L G_{LL} {\Bbb A}_{1,L} \right)
\left( \sum_K G_{KK} {\Bbb A}_{2,K} \right) \ = \ \left( 2{\bf V}_0 
\cdot G \cdot {\Bbb A}_1 \right) \left( 2{\bf V}_0 
\cdot G \cdot {\Bbb A}_2 \right) \ . \cr } $$
We see that exactly because of the GSO condition \GSOcond , $\phi[{\Bbb 
A}_1,{\Bbb A}_2] \eqmodtwo 0$ in this case, meaning that regardless of 
the choice of cocycles the two vertex 
operators commute, in agreement with standard statistics. 

Consider now the case iii), where $S_{{\Bbb A}_1}$ describes a state of 
integer spin and $S_{{\Bbb A}_2}$ one of half odd integer spin.
In this case ${\Bbb A}_{1,L}$ is still integer, whereas the ${\Bbb 
A}_{2,K}$ are half odd integer. We rewrite
$$ {\Bbb A}_2 = ({\Bbb A}_2 - {\bf V}_0) + {\bf V}_0 \ ,  
\nfr{decomp}
where now ${\Bbb A}_2 - {\bf V}_0$ is a vector of integers, and by 
proceeding as before we find
$$ \phi[{\Bbb A}_1,{\Bbb A}_2] \ \eqmodtwo \ ( 2{\bf V}_0 \cdot G \cdot 
{\Bbb A}_1 ) ( 2{\bf V}_0 \cdot G \cdot ({\Bbb A}_2 - {\bf V}_0 )) \ + \ 
2 {\Bbb A}_1 \cdot \widetilde{\bf V}_0 \ ,
\nfr{expressionone}
where we introduced the auxiliary 34-component vector 
(see eq. (2.43) of ref. [\Ref{ammedm}])
$$
(\widetilde{\bf V}_0)_{L}\ \equiv\ \frac12\sum_{K}\widetilde{Y}_{LK}
({\bf V}_0)_{K}\ ,
\efr
which by construction has entries that are $0$ or $1/2$ (mod $1$).
Now, the first term on the right-hand side of \expressionone\ vanishes, 
since $2{\bf V}_0 \cdot G \cdot ({\Bbb A}_2 - {\bf V}_0)$ is manifestly 
integer, and $2{\bf V}_0 \cdot G \cdot {\Bbb A}_1 \eqmodtwo 0$ by the 
GSO condition \GSOcond . But the second term, $2{\Bbb A}_1 \cdot 
\widetilde{\bf V}_0$, can a priori be both an even or an odd integer, 
depending on the choice made for the cocycle matrix and the values taken 
for the ${\Bbb A}_{1,L}$. Thus we do indeed have para-statistics in this 
string model {\it unless} we arrange the cocycle matrix in such a way 
that
$$
\widetilde{\bf V}_0 \eqmodone G \cdot {\bf V}_0 \eqmodone {\bf V}_0 \ .
\nfr{cocycchoice}
Then and only then will the second term on the right-hand side of 
\expressionone\ vanish for all possible values of ${\Bbb A}_{1,L}$,
by virtue of the GSO condition \GSOcond.

It is interesting to note that the cocycle consistency conditions 
described in detail in ref.~[\Ref{ammedm}] allow the vector 
$\widetilde{\bf V}_0$ to take any one of the following four forms
(mod 1),
$$ \left( (1/2)^{34} \right) \qquad
\left( (1/2)^{22} (0)^{12} \right) \qquad
\left( (0)^{22} (1/2)^{12} \right) \qquad
\left( (0)^{34} \right) \ , \efr
but only the first of the four choices 
(i.e. $\widetilde{\bf V}_0 \eqmodone {\bf V}_0 $)
leads to standard statistics 
between physical string states residing in the R and the NS sector, 
whereas the other three choices lead to para-statistics.

We still have to check that a cocycle choice satisfying
\cocycchoice\ also leads 
to standard statistics in the case ii), when both $S_{{\Bbb A}_1}$ and 
$S_{{\Bbb A}_2}$ reside in the R sector. By decomposing both ${\Bbb A}_1$ 
and ${\Bbb A}_2$ according to \decomp\ we find in this case
$$\eqalignno{\phi[{\Bbb A}_1,{\Bbb A}_2] \  \eqmodtwo & \ 
\left( 2 {\bf V}_0 \cdot G \cdot ({\Bbb A}_1 - {\bf V}_0) \right)
\left( 2 {\bf V}_0 \cdot G \cdot ({\Bbb A}_2 - {\bf V}_0) \right)  & 
\nameali{expressiontwo} \cr
& \ + \ {\bf V}_0 \cdot \widetilde{Y} \cdot ({\Bbb A}_2 - {\bf V}_0 ) \
+ \ ({\Bbb A}_1 - {\bf V}_0) \cdot \widetilde{Y} \cdot {\bf V}_0 \ + \
{\bf V}_0 \cdot \widetilde{Y} \cdot {\bf V}_0 \ . \cr } $$
By using the GSO conditions, the first term on the right hand side 
becomes $(1+2k_{00})^2$ mod $2$. The second term we rewrite as
$$\eqalignno{ {\bf V}_0 \cdot \widetilde{Y} \cdot ({\Bbb A}_2 - {\bf 
V}_0)  & \ = \ -({\Bbb A}_2 - {\bf V}_0) \cdot \widetilde{Y} \cdot {\bf 
V}_0 \ + \ 2 {\bf V}_0 \cdot {\rm diag} \widetilde{Y} \cdot ({\Bbb A}_2 
- {\bf V}_0 ) & \numali \cr
&  \eqmodtwo \ -2({\Bbb A}_2 - {\bf V}_0 ) \cdot G \cdot {\bf V}_0 \ - \ 
2\epsilon {\bf V}_0 \cdot G \cdot ({\Bbb A}_2 - {\bf V}_0 ) \
\eqmodtwo \ 0 \ . \cr }  $$
The third term becomes
$$ ({\Bbb A}_1 - {\bf V}_0) \cdot \widetilde{Y} \cdot {\bf V}_0 \ 
\eqmodtwo \ 2({\Bbb A}_1 - {\bf V}_0) \cdot G \cdot
{\bf V}_0 \ \eqmodtwo \ 1+2k_{00} \ ,  
\efr
which cancels the first term mod $2$, since $1+2k_{00}$ is an integer. 
And finally
$$ {\bf V}_0 \cdot \widetilde{Y} \cdot {\bf V}_0 \ = \ {\bf V}_0 \cdot 
{\rm diag} \widetilde{Y} \cdot {\bf V}_0 \ \eqmodtwo \ 1 \ . \efr
Thus, a cocycle choice consistent with eq.~\cocycchoice \ does indeed 
lead to ordinary statistics between any pair of physical vertex 
operators.

One might still wonder whether a cocycle choice exists which satisfies eq.
\cocycchoice . There is no need to worry: A simple analysis
shows that there exist $2^{529}$ such choices!

To conclude, we have seen that already in the simplest conceivable 
four-dimensional KLT model, a
generic choice of cocycles, even one subject to the constraints of 
ref.~[\Ref{ammedm}], gives us vertex operators which satisfy {\sl
para-statistics\/} in the sense of ref. [\Ref{SW}]. But there exists
a class of cocycle choices for which the vertex operators satisfy the
standard statistics.

Since we believe that, once the requirements of ref. [\Ref{ammedm}] are
satisfied, the physics should not depend on the specific choice of cocycles,
we can interpret our result as the proof that, at least in this very 
simple model, the situations leading to para-statistics are 
physically equivalent to one where all vertex operators satisfy the
standard statistics.

The above analysis can be extended to the case of general four-dimensional
heterotic KLT string models, but a full discussion is technically 
quite complicated and will be reported in a future publication.
\chapter{Conclusions and outlook}
In this final section we would like to make a few comments on what we have
seen so far.

We have introduced a world-sheet transformation applicable to
any string theory in a $D$ dimensional ($D$ even) Minkowski
background which gives rise to the space-time CPT transformation when
acting on vertex operators of physical string states.
We have also shown that under the
hypothesis of Lorentz invariance and spin-statistics
the CPT theorem holds, at least to any order in perturbation theory. 

That being the case, any breaking of CPT invariance in such a string 
theory would have to be due either to subtle
non-perturbative effects, like those proposed by Ellis et 
al.~[\Ref{Ellis}], or due to a spontaneous move away from the original 
background, in the manner of the mechanism suggested by Kostelecky and 
Potting~[\Ref{Pott}], where some field in the low-energy effective field 
theory which is odd under CPT (for example a vector field) 
is imagined to acquire a nonzero vacuum expectation value.
This mechanism naturally leads us to consider to which extent it 
is possible to generalize the CPT theorem to general backgrounds.
The answer to this question might be more easily perceived if we could 
find a proof of the CPT theorem which does not use so directly the 
hypothesis of Lorentz invariance, but relies more on just the 
world-sheet properties of the strings. 

The relation between space-time spin and world-sheet statistics of 
physical state vertex operators is another point that deserves a more 
thorough investigation.
This obviously brings up the issue of what are the properties satisfied by a
string theory in analogy with the Wightman axioms in field theory, or in
other words, which are the {\it minimum} assumptions that we can make in 
trying to define and prove the spin-statistics theorem or the CPT theorem 
in string theory. 

It is clear that a full understanding of the role of CPT invariance and 
spin-statistics in string theory is still lacking. 
Perhaps we will have to await an improved understanding of the nature 
of string theory itself.

\appendix{Helicity in String Theory}
\setchap{helicitychap}
In this appendix we give our conventions for helicity and show in detail 
how the helicity sign change under CPT comes about in an explicit 
example.

We define the helicity as the inner product of the angular momentum 
$\vec{J}$ with a unit vector pointing in the direction of the momentum,
$$ \eta \ = \ { \vec{J} \cdot \vec{k} \over |\vec{k}|} \ = \
{1 \over 2} \epsilon_{ijk} M^{jk} k^i {1 \over |\vec{k}|} \ ,
\nfr{heldef}
where the generators $M^{\mu \nu}$ of the Lorentz group are given by
$$\eqalignno{ M^{\mu \nu} \ = & \oint_0 {{\rm d} z \over 2\pi i} \left[
{-1 \over 2i} \left( X^{\mu} \partial_z X^{\nu} - X^{\nu} \partial_z 
X^{\mu} \right) -i \psi^{\mu} \psi^{\nu} \right] & \nameali{lorentzgen} 
\cr
& + \ \oint_0 {{\rm d} \bar{z} \over 2\pi i} \left[
{-1 \over 2i} \left( X^{\mu} \partial_{\bar{z}} X^{\nu} - X^{\nu} 
\partial_{\bar{z}} 
X^{\mu} \right) \right] \ . \cr } $$
In section 3 we argued on general grounds that the helicity is flipped
by the space-time CPT transformation. As an example, consider the 
physical vertex operator \groundvert\ in the case where this describes a
string state of spin $1/2$. Acting on a space-time spinorial state $\vert 
k,{\Bbb A}\rangle$ created from the conformal vacuum by means of the 
operator \groundst ,
$$ M^{\mu \nu} \ \vert k , {\Bbb A} \rangle \ = \
\left\{ (q^{\mu} k^{\nu} - q^{\nu} k^{\mu} ) \delta_{\Bbb A}^{\ {\Bbb 
B}} \ + \ {i \over 4} \left[ {\bfmath \Gamma}^{\mu}, {\bfmath 
\Gamma}^{\nu} \right]_{\Bbb A}^{\ {\Bbb B}} \right\} \ \vert k, {\Bbb B} 
\rangle \ , \efr
where $q^{\mu}$ is the operator conjugate to $k^{\nu}$ and the gamma 
matrices are defined by eq.~\gammadef . 

Therefore, in order for the physical state created by the vertex 
operator \groundvert\ to have helicity $\eta$, the space-time spinor
$W^{\Bbb A}_{(in)} (k,\eta,\{\lambda\})$ should satisfy the matrix 
eigenvalue equation
$$ \left(W_{(in)}(k,\eta,\{\lambda\}) \right)^T \left( {i \over 8} 
\epsilon_{ijk} {k^i \over \vert \vec{k} \vert} 
\left[ {\bfmath \Gamma}^{j}, {\bfmath \Gamma}^{k} \right] - \eta 
\right) \ = \ 0 \ . \nfr{helone}
Assuming this to be the case we would like to verify that the CPT 
transformed vertex operator \spincpt\ has helicity $-\eta$, i.e. that
$$ \left(W_{(in)}(k,\eta,\{\lambda\}) \right)^{\dagger} {\bfmath 
\Gamma}^0 {\Bbb C}^{T,-1} {\bfmath \Gamma}_5 \ \left( {i \over 8} 
\epsilon_{ijk} {k^i \over \vert \vec{k} \vert } 
\left[ {\bfmath \Gamma}^{j}, {\bfmath \Gamma}^{k} \right] + \eta 
\right) \ = \ 0 \ . \nfr{heltwo}
This follows immediately if we take the complex conjugate of 
eq.~\helone\ and multiply from the right by ${\bfmath 
\Gamma}^0 {\Bbb C}^{T,-1} {\bfmath \Gamma}^5$, using the properties
$$\eqalignno{ & {\bfmath \Gamma}^0 {\bfmath \Gamma}^{\mu} \ = \ - \left( 
{\bfmath \Gamma}^{\mu} \right)^{\dagger} {\bfmath \Gamma}^0 \ , & 
\nameali{relations} \cr
& \left\{ {\bfmath \Gamma}^5 , {\bfmath \Gamma}^{\mu} \right\} \ = \  0 
\ , \cr
& {\bfmath \Gamma}^{\mu} {\bfmath \Gamma}^{\nu} {\Bbb C} \ = \
{\Bbb C} \left( {\bfmath \Gamma}^{\mu} \right)^T 
\left( {\bfmath \Gamma}^{\nu} \right)^T \ . \cr} $$
\references
\beginref  
\Rref{KLT}{H.~Kawai, D.C.~Lewellen and S.-H.H.~Tye, Nucl.Phys. 
{\bf B288} (1987) 1;\newline
H.~Kawai, D.C.~Lewellen, J.A.~Schwartz and S.-H.H.~Tye, Nucl.Phys. 
{\bf B299} (1988) 431.}
\Rref{Anto}{I.~Antoniadis, C.~Bachas, C.~Kounnas and P.~Windey, 
Phys.Lett. {\bf 171B} (1986) 51; \newline
I.~Antoniadis, C.~Bachas and C.~Kounnas, Nucl.Phys. {\bf B289} (1987) 87;
\newline I.~Antoniadis and C.~Bachas, Nucl.Phys. {\bf B298} (1988) 586.}
\Rref{Pott}{V.A.~Kostelecky and R.~Potting, Nucl.Phys. {\bf B359} (1991)
545; \newline preprint \hbox{hep-ph/9211116}.}
\Rref{Sonoda}{H.~Sonoda, Nucl.Phys. {\bf B326} (1989) 135.}
\Rref{Weisberger}{J.L.~Montag and W.I.~Weisberger, Nucl.Phys. {\bf B363}
(1991) 527.}
\Rref{hokgid}{E.~D'Hoker and S.B.~Giddings, Nucl.Phys. {\bf B291} (1987) 90.}
\Rref{Berera}{A.~Berera, Nucl.Phys. {\bf B411} (1994) 157.}
\Rref{Hoker}{K.~Aoki, E.~D'Hoker and D.H.~Phong, Nucl.Phys. {\bf B342}
(1990) 149;\newline
E.~D'Hoker and D.H.~Phong, Phys.Rev.Lett. {\bf 70} (1993) 3692; \newline 
preprint \hbox{hep-th/9404128};\newline Nucl.Phys. {\bf B440} (1995) 24, 
\hbox{hep-th/9410152}.}
\Rref{GSW}{M.B.~Green, J.H.~Schwarz and E.~Witten, ``{\sl Superstring 
Theory},'' Cambridge University Press, 1987.}
\Rref{Bluhm}{R.~Bluhm, L.~Dolan and P.~Goddard, Nucl.Phys. {\bf B309}
(1988) 330.}
\Rref{IZ}{C.~Itzykson and J.-B.~Zuber, ``{\sl Quantum Field Theory},"
McGraw-Hill, New York, 1980.}
\Rref{ammedm}{A.~Pasquinucci and K.~Roland, Nucl.Phys. {\bf 440} (1995) 441, 
\hbox{hep-th/9411015}.}
\Rref{mink}{A.~Pasquinucci and K.~Roland, Phys.Lett. {\bf B351} (1995) 131, 
\hbox{hep-th/9503040}.}
\Rref{normaliz}{A.~Pasquinucci and K.~Roland, Nucl.Phys. {\bf B457} (1995) 
27, \hbox{hep-th/9508135}.}
\Rref{Weinbook}{S.~Weinberg, ``{\sl The Quantum Theory of Fields\/}", 
Cambridge University Press, 1995. }
\Rref{SW}{R.F.~Streater and A.S.~Wightman, ``{\sl PCT, Spin and Statistics,
and all that\/},'' Benjamin Inc., New York, 1964.}
\Rref{Luders}{G.~L\"uders, Ann.Phys. {\bf 2} (1957) 1.}
\Rref{Ellis}{J.~Ellis, N.E.~Mavromatos and D.V.~Nanopoulos, Phys.Lett.
{\bf B293} (1992) 37;\newline
CPLEAR Collaboration, J.~Ellis, J.L.~Lopez, N.E.~Mavromatos and 
D.V.~Nanopoulos, ``{\sl Tests of CPT symmetry and quantum mechanics with 
experimental data from CPLEAR\/}", preprint \hbox{hep-ex/9511001}.}
\Rref{Pauli}{W.~Pauli, ``{\sl Niels Bohr and the Development of Physics\/}", 
McGraw-Hill, New York, 1955;\newline
NuovoCimento {\bf 6} (1957) 204.}
\Rref{Wittwcpt}{E.~Witten, Comm.Math.Phys. {\bf 109} (1987) 525.}
\Rref{uswcpt}{A. Pasquinucci and K. Roland, to appear.}
\Rref{martcaus}{E.~Martinec, preprint \hbox{hep-th/9311129}.}
\Rref{elienonloc}{D.A.~Eliezer and R.P.~Woodard, Nucl.Phys. {\bf B325} 
(1989) 389.}
\Rref{lowenonloc}{D.A.~Lowe, L.~Susskind and J.~Uglum, Phys.Lett {\bf B327}
(1994) 226, \hbox{hep-th/9402136}.}
\Rref{Hawking}{S.~Hawking, Phys.Rev.{\bf D32} (1985) 2489.}
\Rref{others}{J.~Schwinger, Phys.Rev.{\bf 82} (1951) 914; \newline
J.S.~Bell, Proc.Roy.Soc. (London) {\bf A231} (1955) 479; \newline
G.~L\"{u}ders and B.~Zumino, Phys.Rev. {\bf 106} (1957) 345; \newline
R.~Jost, ``{\it The General Theory of Quantized Fields},'' AMS, Providence, 
1965.}
\Rref{BPZ}{A.A.~Belavin, A.M.~Polyakov and A.B.~Zamolodchikov, 
Nucl.Phys. {\bf B241} (1984) 333.}
\Rref{MooreSeiberg}{G.~Moore and N.~Seiberg, Comm.Math.Phys. {\bf 123} 
(1989) 177.}
\Rref{seiberg-witten}{N.~Seiberg and E.~Witten, Nucl.Phys. {\bf B276} 
(1986) 272.}
\Rref{DHR}{S.~Doplicher, R.~Haag and J.E.~Roberts, Comm.Math.Phys. 
{\bf 23} (1971) 199, {\bf 35} (1974) 49.}
\Rref{Klein}{O.~Klein, J.Phys.Radium {\bf 9} (1938) 1;\newline
G.~L\"uders, Z.Naturforsch. {\bf 13a} (1958) 254;\newline
P.~Jordan and E.~Wigner, Z.Physik {\bf 47} (1928) 631;\newline
H.~Araki, J.Math.Phys {\bf 2} (1961) 267;\newline
R.~Haag, Dan.Mat.Fys.Medd. {\bf 29} (1955) 12.}
\Rref{GSO}{F.~Gliozzi, J.~Sherk and D.I.~Olive, Phys.Lett. {\bf 65B} 
(1976) 282, Nucl.Phys. {\bf B122} (1977) 253.}
\Rref{otherSS}{M.~Fierz, Helv.Phys.Acta {\bf 12} (1939) 3;\newline
W.~Pauli, Phys.Rev. {\bf 58} (1940) 716;\newline
G.~L\"uders and B.~Zumino, Phys.Rev. {\bf 110} (1958) 1450;\newline
N.~Burgoyne, NuovoCimento {\bf 8} (1958) 607.}
\Rref{parastat}{Y.~Ohnuki and S.~Kamefuchi, Phys.Rev. {\bf 170} (1968) 
1279, Ann.Phys. {\bf 51} (1969) 337;\newline
K.~Dr\"{u}hl, R.~Haag and J.E.~Roberts, Comm.Math.Phys. {\bf 18} (1970) 
204.}
\endref
\ciao
